\documentclass[journal]{IEEEtran}

\usepackage{amsmath,amsfonts}

\usepackage[dvipsnames]{xcolor}

\usepackage{cite}

\usepackage{graphicx}

\usepackage{svg}

\usepackage{amssymb}

\usepackage{multirow}
\usepackage{balance}
\usepackage{array}
\usepackage{makecell}

\usepackage{hyperref}

\ifCLASSINFOpdf
\else
\fi
\begin{document}
%
\title{The Uniqueness Theorem for Nonlocal
Hydrodynamic Media}
%
%
%

\author{Christos~Mystilidis,~\IEEEmembership{Graduate~Student~Member,~IEEE,}
         George~Fikioris,~\IEEEmembership{Senior~Member,~IEEE,}
         Christos~Tserkezis,
         Guy~A.~E.~Vandenbosch,~\IEEEmembership{Fellow,~IEEE,}~and
         Xuezhi~Zheng,~\IEEEmembership{Member,~IEEE.}
\thanks{ C. Mystilidis, G. A. E. Vandenbosch, and X. Zheng are with the Department of Electrical Engineering (ESAT-WaveCore), KU Leuven, B-3001 Leuven,
Belgium (email: xuezhi.zheng@esat.kuleuven.be)\emph{(Corresponding author: Xuezhi Zheng.) }.

G. Fikioris is with the School of Electrical and Computer Engineering, National Technical University of Athens, 157 80 Athens, Greece.

C. Tserkezis is with POLIMA: Center for Polariton-driven Light--Matter Interactions, University of
Southern Denmark, DK-5230 Odense, Denmark.

CM, GAEV, and XZ were supported by the KU Leuven internal funds: the C1 Project No. C14/19/083, the IDN Project No. IDN/20/014, and the small infrastructure Grant No. KA/20/019. 

CM acknowledges funding from IEEE Antennas \& Propagation Society (AP--S) through the Doctoral Research Grant 2022. 

CT acknowledges funding from VILLUM Fonden (Grant No. 16498).
POLIMA is sponsored by the Danish National Research
Foundation (Project No. DNRF165).
}}

\markboth{}%
{Shell \MakeLowercase{\textit{et al.}}: Bare Demo of IEEEtran.cls for IEEE Journals}

%



\maketitle

\begin{abstract}
We investigate a fundamental electromagnetic theorem, namely the uniqueness theorem, in the context of nonlocal electromagnetics, as simulated by a popular semiclassical model, the Hydrodynamic Drude Model (HDM) and extensions thereof such as the Generalized Nonlocal Optical Response (GNOR). The derivations and proofs presented here give a theoretical foundation to the use of the Additional Boundary Conditions (ABCs), whose necessity is recognized and underlined in virtually all implementations and applications of HDM. Our proofs follow a mathematically relaxed style, borrowing from the literature of established electromagnetics textbooks that study the matter from an engineering perspective. Through this simpler route we deduce clear and intuitive material-response requirements for uniqueness to hold, while using a familiar parlance in a topic that is mostly studied through a physics perspective. Two numerical examples that examine the problem from either a semianalytical or a purely numerical viewpoint support our findings.
\end{abstract}


%
\IEEEpeerreviewmaketitle

\section{Introduction}
\label{sec:intro}
\IEEEPARstart{T}{he} \emph{uniqueness theorem} is one of the most powerful and  physically intuitive results in \emph{macroscopic electromagnetics}. Indeed, a physical problem with given sources should accept a single physical solution, and the mathematical formulation of said problem must abide by this principle. But further, the uniqueness theorem constitutes a stepping stone for all solution strategies in macroscopic electromagnetics, especially heuristic ones: no matter how \emph{a} solution is found, it is, by virtue of the uniqueness theorem, \emph{the} solution of the problem. Such considerations are made (often implicitly) when we use the equivalence principle, Huygens's principle, the image theorem, and the induction theorem, to name just a few \cite{kong}.

The proof of the uniqueness theorem for the most ``well-behaving'' media, is included in standard and excellent electromagnetics textbooks, examining the subject either from a physicist's or an engineer's perspective \cite{kong,stratton,collin,balanis}. The proof is both straightforward and instructive; it allows students to better digest the necessity of boundary (and initial) conditions and, further, the impact that \emph{constitutive relations} have on the mathematical complexity of the problem; and it provides specialists a directly applicable recipe for treating active-research problems.

For complex (i.e., anisotropic, inhomogeneous, nonlocal, etc.) materials, which may promise much more attractive engineering applications, the constitutive relations become complicated (e.g., position-dependent, tensorial, etc.). This results in nontrivial and nonstandard extensions of the uniqueness theorem. The proof is then the subject of research works and advanced textbooks \cite{chew}. For example, non-chiral, bi-isotropic media (Tellegen media) have been discussed in \cite{piers}, inhomogeneous bi-anisotropic media in \cite{bian}, and lossy, anisotropic, inhomogeneous media with diagonal material tensors appropriate for invisibility-cloaks engineering are given in \cite{reza}.

But even standard media, such as simple and noble metals, may exhibit a complex material response. For metallic nanostructures, when the characteristic length scale (e.g., the vanishing gap in the Nanoparticle on Mirror--NPoM--configuration \cite{baumberg_natmat18}) becomes comparable to the \emph{nonlocal length scale}, then corrections to the standard macroscopic electromagnetics are anticipated \cite{mortensen_rev}. \emph{Nonlocality} implies that the response at a given point of the material is determined by a large number of individual microscopic interactions over a volume surrounding this very point (and demarcated by the aforesaid nonlocal length scale) \cite{micro-macro}. This phenomenon of microscopic origins requires, in principle, a full microscopic theory to be properly accounted for. Nonetheless, \emph{semi-classical models}, which attempt to combine the sturdy framework of Maxwell's equations with the desired material response have enjoyed much popularity \cite{chapter}. In particular, the \emph{Hydrodynamic Drude Model} (HDM), which provides an extension to the microscopic Ohm's law \cite{phenomen}, experienced a revitalization in the past decade, by virtue of its ability to predict accurately the near-field enhancement in NPoM structures \cite{ciraci_science} and the size-dependent blueshifting of the scattering spectrum in electron energy loss spectroscopy and far-field spectroscopy experiments involving noble metals \cite{eels,shen}, combined with a simple, numerically and analytically amenable framework. Today, it enjoys widespread applications \cite{nonlin,yoo,uni,zouros,filipa,hanson}, and powerful extensions \cite{gnor,schdm,lqht}. Among them, the Generalized Nonlocal Optical Response (GNOR) is particularly simple and accurate (within its range of validity). It introduces a classical diffusive term in the equation of motion of free electrons, which, as more recent perspectives suggest, incorporates (effectively) an additional quantum phenomenon, namely Landau damping, in the optical response \cite{gnor_per,newgnor}. It is interesting that, aside from metals, different incarnations of nonlocality have been studied in parallel through the lens of HDM and its variants in metamaterials \cite{hyper,meta1,meta3}, graphene \cite{graphene1,graphene2}, and in polar dielectrics \cite{dnl}.

The theoretical and experimental interest in nonlocality and the HDM is accompanied by the presentation of various solution strategies, which are well-known in macroscopic electromagnetics and microwave frequencies. One could list semi-analytical approaches that use Transformation Optics and $S$ matrices \cite{pendry,bene,dong}, as well as full-blown numerical algorithms like the Finite Element Method \cite{modified,fem}, the discontinuous Galerkin method \cite{li,hdg}, the Boundary Element Method (BEM) \cite{mort-bem,hoh,bem1,bem2,bem3,bem4,bem2d}, the Finite Difference Time Domain method \cite{schatz,fdtd}, the Discrete Sources Method \cite{yuri3,yuri1,yuri2}, and the Volumetric Method of Moments (MoM) \cite{chen23}. All of these methods invoke, implicitly or explicitly, the uniqueness theorem.

The question of existence and uniqueness of a solution to the coupled system of Maxwell's equations and HDM has been previously addressed in the mathematical literature. \cite{math2} deals with the existence and uniqueness of a solution to the weak formulation of the coupled system of the wave equation of the electric field and the HDM in the frequency domain. In \cite{math1,math3} the existence and uniqueness of the system of Maxwell's equations with the HDM and the continuity equation, in time domain, is discussed in detail. Though \cite{math1} uses rather standard boundary conditions, \cite{math3} enriches the study by adding other sets, named ``electric'' and ``magnetic'' boundary conditions. 

In this work, we present a proof of the uniqueness theorem for the coupled system of Maxwell's equations and the HDM with either real or complex hydrodynamic parameter in the frequency domain. Importantly, we use the \emph{Additional Boundary Conditions} (ABCs) which traditionally accompany the HDM. We focus on the set of ABCs that is most typically used by the nanoplasmonics community. We stress that the proofs presented herein follow deliberately a mathematically relaxed style. For example, we do not pay attention to the often complicated question of mathematical spaces to which the solution domain $\Omega$, the boundary data, and the solutions $\mathbf{E}$, $\mathbf{H}$ are supposed to belong, Thus, the present methods, based on simple calculus and adaptations to the standard procedure that is presented in established engineering textbooks such as \cite{kong,chew}, concentrate on aspects of uniqueness that the applied electromagnetics community generally considers more salient. What this approach undeniably lacks in mathematical rigor (see for example \cite{kress} and the aforementioned mathematical papers) is recovered from its educational approach, especially concerning the necessary modifications with respect to the familiar proofs from macroscopic electromagnetics. In particular, the influence of the involved HDM on the material-response requirements figures prominently in this work. 

After a quick recapitulation of key notions of the HDM in Section \ref{sec:hdm}, we proceed with the proof in Section \ref{sec:der}. An Extension of the uniqueness theorem is discussed in Section \ref{sec:ext}. Numerical experiments in Section \ref{sec:num} support the results. We assume ---and subsequently suppress--- a $\exp{(-i\omega t)}$ time dependence. SI units are used throughout. 

\section{The Hydrodynamic Drude Model}
\label{sec:hdm}
Mathematically, the HDM can be summarized in a single material equation \cite{micro-macro}
\begin{equation}
    \label{eq:hdm}
    \frac{\beta^2}{\omega(\omega+i\gamma)} \nabla \left(\nabla \cdot \mathbf{P}_f(\mathbf{r})) + \mathbf{P}_f(\mathbf{r}\right) = -\epsilon_0 \frac{ \omega_p^2}{\omega(\omega+i\gamma)} \mathbf{E}(\mathbf{r}). 
\end{equation}
Above, $\epsilon_0$ is the vacuum permittivity, $\mathbf{r}$ is a spatial point within the nonlocal medium, $\gamma$ is a phenomenological damping rate which describes all energy losses from the system of fields and collective electron motion \cite{forstmann}, $\omega_p$ is the plasma frequency, and $\mathbf{E}$ is the electric field vector. $\mathbf{P}_f$ denotes the free electron polarization density. Its partial derivative with respect to time is the current density of free electrons, that is, $\mathbf{J}_f=-i\omega\mathbf{P}_f$ \cite{micro-macro}. Finally, $\beta$ is the hydrodynamic parameter, typically taken equal to the high-frequency limit $\sqrt{3/5}\upsilon_F$, when $\omega \gg \gamma$ \cite{raza}; $\upsilon_F$ signifies the Fermi velocity. In the limit $\beta \rightarrow 0$, the \emph{local} response is retrieved; and if one further introduces $\mathbf{J}_f$ in (\ref{eq:hdm}), the standard Ohm's law is obtained ($\mathbf{J}_f(\mathbf{r}) = \sigma(\omega) \mathbf{E}(\mathbf{r})$, where $\sigma$ is the standard AC Drude conductivity) \cite{modified}. To analyze a system within the HDM, (\ref{eq:hdm}) must be coupled to the familiar Maxwell's equations of macroscopic electromagnetics.

The new constitutive relation, whose nonlocal character is evident from the spatial derivatives, addresses exclusively free electrons. We note that an alternative interaction mechanism between light and electrons is provided by the bound electrons. This is a purely local interaction; we neglect it in what follows (see Section \ref{sec:der}).
\begin{figure}[h!]
    \centering
    \includegraphics[scale=0.5]{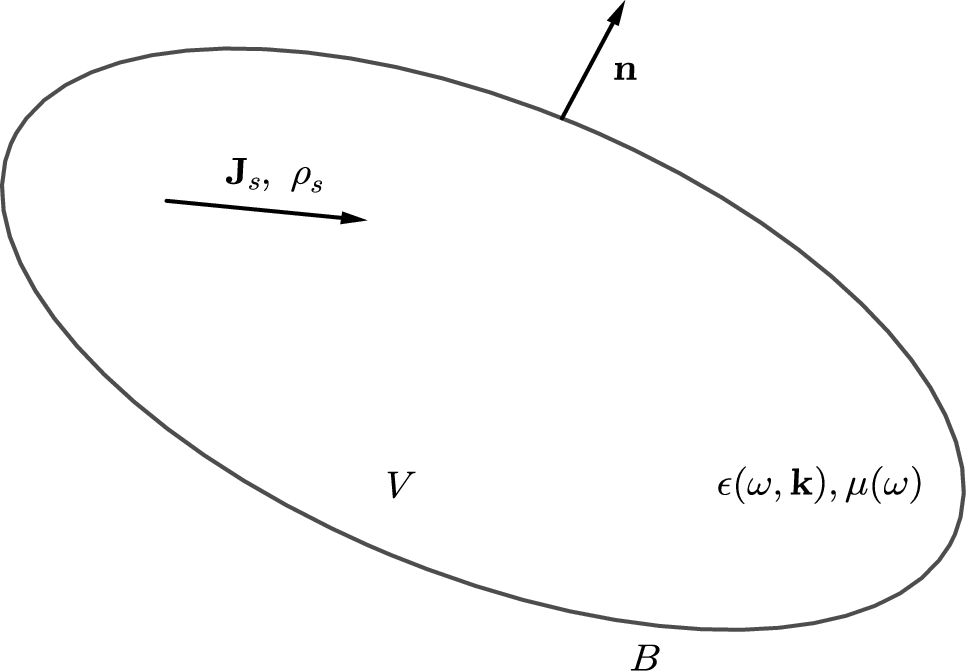}
    \caption{\emph{The geometry of the problem}. We demonstrate the uniqueness theorem for an arbitrary object, occupying a volume $V$, which is enclosed  by the boundary $B$. The vector unit normal to the surface is $\mathbf{n}$ and points from the interior to the exterior. Current and charge sources $\mathbf{J}_s$ and $\rho_s$ may lie inside it. Its material composition is described by the magnetic permeability $\mu(\omega)$, and the nonlocal, within the HDM, electric permittivity $\epsilon(\omega,\mathbf{k})$.}
    \label{fig:geometry}
\end{figure}
The introduction of a radically different constitutive relation as (\ref{eq:hdm}) modifies the standard arguments for uniqueness of the solution of Maxwell's equations. To have a solution that is unique, the second-order differential operator requires the imposition of ABCs. 

A more physically intuitive way to realize the necessity of introducing ABCs is by examining Maxwell's equations in a generic \emph{spatially dispersive} medium. It can be easily shown, that the wavevector-dependence of the electric permittivity allows for \emph{longitudinal waves} to be solutions of the homogeneous Maxwell equations \cite{maier}. These longitudinal waves are \emph{uncoupled} with the traditionally expected transverse waves, save for the interfaces \cite{raza}. As such, they constitute a new degree of freedom that must be constrained by appropriate boundary conditions, hence the necessity of ABCs. Within the HDM, longitudinal fields are driven by an electron gas (within Thomas-Fermi kinetics) pressure term \cite{forstmann}, which is intertwined to the differential operator correction in (\ref{eq:hdm}). 

During the 70s and 80s, the ABCs were a subject of discussion (sometimes with a level of contention) \cite{sauter,melnyk,fs,br,forstmann}. Often, the arguments were  heuristic, as the ABCs were judged by comparing to experiments. The contemporary perspective argues that the choice of the ABCs is not a matter of debate; the proper ABCs follow necessarily from the  physical assumptions, i.e., from the governing equation (\ref{eq:hdm})  \cite{mortensen_rev}. This argument has persuasively led to the \emph{Sauter ABC} \cite{sauter}, which stipulates that the normal component of the current $\mathbf{J}_f$ (or $\mathbf{P}_f$) vanishes at the interface, for the case of a nonlocal--local (metal-dielectric) interface, and which arises from the assumption that the equilibrium electron density is constant inside the volume of the nonlocal medium and vanishes abruptly on its geometrical surface and beyond \cite{raza}. However, the situation concerning the nonlocal--nonlocal interface (an interface between two metals) appears cloudy \cite{stam}. 

\section{The Uniqueness Theorem for Simple Metals}
\label{sec:der}

We present results for the HDM when applied to metals that possess no bound electrons. We stress that the HDM has consistently given underwhelming results for such materials \cite{tepe,stella,schdm}; its success hinges upon the suppression of the \emph{electron spill-out}, an assumption inherent to the selected (as above) equilibrium electron density. Electron spill-out may be indeed negligible when the metal possesses a sufficiently high work function; this is not the case for simple (but it is quite accurate for noble) metals \cite{schdm,stam}. Further, in theoretical works such as \cite{modified,unusual}, HDM neglecting interband transitions was applied even for metals that demonstrate significant contribution by bound electrons:  On the one hand, the extension to include interband transitions is straightforward. On the other, this version of the HDM captures salient features of the nonlocal response, especially the existence of additional longitudinal resonances above the plasma frequency; since bound electrons dominate the high frequency spectrum \cite{olmon}, longitudinal resonances tend to be sidelined when the bound-electron contribution is included. As in said works, focusing on the free-electron contribution allows us to demonstrate clearly the complications introduced by the HDM \emph{per se}. The inclusion of bound electrons in the response will be the subject of future work; still, in the next Section we introduce additional complexity, which increases HDM's predictive power, but which blurs in tandem the derivation with algebraic details. This is achieved by substituting the real $\beta^2$ in this Section, by a complex one in the next.

The derivations concern the geometry of  Fig. \ref{fig:geometry}. The nonlocal material occupies an arbitrary volume $V$ in space and is limited by the boundary $B$. Inside the medium there may be imposed current and charge sources $\mathbf{J}_s$ and $\rho_s$. The unit normal $\mathbf{n}$ on the surface is directed from the inside to the outside of the scatterer. The background, though inconsequential to the derivation, will be assumed nonlocal, the most complex case; we will explicitly discuss the nonlocal--local interface shortly.

Following the standard procedure \cite{kong}, we assume the existence of two solutions of Maxwell's equations and the hydrodynamic equation, to be denoted by $\mathbf{E}_1$,  $\mathbf{H}_1$ and $\mathbf{E}_2$, $ \mathbf{H}_2$, where
$\mathbf{H}$ is the magnetic field. These two solutions satisfy
\begin{equation}
        \label{eq:faraday}
        \nabla \times \mathbf{E}_{1,2} = i\omega\mu\mathbf{H}_{1,2},
    \end{equation}
\begin{equation}
        \label{eq:ampere}
        \nabla \times \mathbf{H}_{1,2} = -i\omega\mathbf{D}_{1,2} + \mathbf{J}_s,
    \end{equation}
as well as (\ref{eq:hdm}). We solve this coupled system assuming, initially, that \emph{all} the following boundary conditions hold: 
 (a) the tangential components of the electric and the magnetic field are given on the boundary
\begin{equation}
    \label{eq:bound1}
    \mathbf{n} \times \mathbf{E}_{1,2} = \mathbf{f}(\mathbf{r}) \quad \text{or} \quad \mathbf{n} \times \mathbf{H}_{1,2} = \mathbf{h}(\mathbf{r}),
\end{equation}
where $\mathbf{f}$ and $\mathbf{h}$ are vector functions and with $\mathbf{r}$ belonging to $B$ or to a subset of $B$, $B_1$ and $B_2$ respectively, such that \mbox{$B_1 \cup B_2 = B$}. When both conditions are specified over the same spatial points of $B$ (or a subset thereof), then the conditions (namely $\mathbf{f}$ and $\mathbf{h}$) must be \emph{compatible} \cite{kong}.  (b) The normal component of the free-electron polarization density $\mathbf{P}_f$ is given on the boundary
\begin{equation}
    \label{eq:bound2}
    \mathbf{n} \cdot \mathbf{P}_{f,1,2} = p(\mathbf{r}).
\end{equation}
The ABC above is a generalization of the aforementioned Sauter ABC \cite{forstmann}. (c) A quantity involving the divergence of the electric field is given on the boundary
\begin{equation}
    \label{eq:bound3}
    \frac{\beta^2}{\omega_p^2} \nabla \cdot \mathbf{E}_{1,2} = s(\mathbf{r}).
\end{equation}
The equation corresponds to the \emph{Forstmann-Stenschke ABC} and stems from the requirement of continuous normal component of the energy current density \cite{fs}. The two ABCs are imposed in the same manner as the standard ones, applied on the whole boundary or a part of it (and with compatibility, referring to $p$ and $s$, still required, if necessary).

Before we continue, it deserves to be mentioned that the ABC
\begin{equation}
    \label{eq:bound2_alt}
    \mathbf{n} \cdot \mathbf{E}_{1,2} = u(\mathbf{r}),
\end{equation}
when bound electrons are neglected, is implied by (\ref{eq:bound2}) \cite{hyper}. In particular, for the case of the metal--dielectric interface and including bound electrons, it is shown in \cite[Appendix,~comment~1]{yuri3} that (\ref{eq:bound2_alt}) (and of course (\ref{eq:bound2})) arise from the assumption of a step profile of the ground electron density, inherent and central to the HDM. Now, since (\ref{eq:bound2_alt}) involves the field, it is more convenient and thus used in several papers \cite{hyper,waveguide,mort-bem,yuri3}. Equation (\ref{eq:bound2_alt}) arises by combining the (standard) Maxwell boundary condition of continuity of the electric displacement (in the absence of free charges) and the discussed generalization of the Sauter ABC \cite{waveguide}. The field format will be used in the derivations below.

In (\ref{eq:faraday}) we assumed the constitutive relation 
\mbox{$\mathbf{B}_{1,2}=\mu \mathbf{H}_{1,2}$}. $\mathbf{B}$ is the magnetic induction field, and $\mu$ is the magnetic permeability of the medium; at optical frequencies it assumes the vacuum value $\mu_0$ for natural media \cite{magn}, but we will treat it here as a complex function of frequency. In the absence of bound-electron contributions, $\mathbf{D}$ is written $\mathbf{D}_{1,2} = \epsilon_0\mathbf{E}_{1,2} + \mathbf{P}_{f,1,2}$. Eliminating $\mathbf{P}_{f,1,2}$ from (\ref{eq:ampere}) (see also \cite{gen})
\begin{equation}
    \label{eq:pf}
    \mathbf{P}_{f,1,2} = - \epsilon_0\frac{\omega_p^2}{\omega(\omega + i\gamma)} \left[ \mathbf{E}_{1,2} - \frac{\beta^2}{\omega_p^2} \left( \nabla (\nabla \cdot \mathbf{E}_{1,2}) - \frac{\nabla \rho_s}{\epsilon_0}\right) \right],
\end{equation}
and substituting in (\ref{eq:ampere})
\begin{equation}
    \begin{split}
        \label{eq:ampere_pelim}
        \nabla \times \mathbf{H}_{1,2} =& -i\omega \bigg(\epsilon \mathbf{E}_{1,2} + \\ &\epsilon_0\frac{\beta^2}{\omega(\omega + i\gamma)} \left[ \nabla(\nabla \cdot \mathbf{E}_{1,2}) -\frac{\nabla\rho_s}{\epsilon_0} \right]\bigg) + \mathbf{J}_s.
    \end{split}
\end{equation}
Above, $\epsilon = \epsilon_0(1+\chi_f)$ is the \emph{transverse} electric permittivity; $\chi_f(\omega) = -\omega^2_p/(\omega^2+i\omega\gamma)$ is the Drude free-electron susceptibility.

Next, we take the \emph{difference fields} $\overline{\mathbf{H}}=\mathbf{H}_1 - \mathbf{H}_2$ and \mbox{$\overline{\mathbf{E}}=\mathbf{E}_1 - \mathbf{E}_2$}. The sources are eliminated, so that  $\overline{\mathbf{E}}$ and $\overline{\mathbf{H}}$ satisfy 
\begin{equation}
        \label{eq:faraday_bar}
        \nabla \times \overline{\mathbf{E}} = i\omega\mu\overline{\mathbf{H}},
\end{equation}
\begin{equation}
    \label{eq:ampere_bar}
    \nabla \times \overline{\mathbf{H}} = -i\omega \left(\epsilon \overline{\mathbf{E}} + \epsilon_0\frac{\beta^2}{\omega(\omega + i\gamma)} \nabla(\nabla \cdot \overline{\mathbf{E}})  \right),
\end{equation}
as well as  the homogeneous versions of (\ref{eq:bound1}), (\ref{eq:bound3}), and (\ref{eq:bound2_alt}). We take then the complex conjugate (denoted by $*$) of (\ref{eq:faraday_bar}) and multiply it by $\overline{\mathbf{H}}$, and multiply (\ref{eq:ampere_bar}) by $\overline{\mathbf{E}}^*$, leading to
\begin{equation}
        \label{eq:faraday_conj}
        \overline{\mathbf{H}} \cdot \nabla \times \overline{\mathbf{E}}^* = -i\omega\mu^* \overline{\mathbf{H}} \cdot \overline{\mathbf{H}}^*,
\end{equation}
\begin{equation}
    \label{eq:ampere_conj}
    \overline{\mathbf{E}}^* \cdot \nabla \times \overline{\mathbf{H}} = -i\omega \overline{\mathbf{E}}^* \cdot \left(\epsilon \overline{\mathbf{E}} + \epsilon_0\frac{\beta^2}{\omega(\omega + i\gamma)} \nabla(\nabla \cdot \overline{\mathbf{E}})  \right).
\end{equation}
Now subtract (\ref{eq:ampere_conj}) from (\ref{eq:faraday_conj}) and use the vector identity $\overline{\mathbf{H}} \cdot \nabla \times \overline{\mathbf{E}}^{*} - \overline{\mathbf{E}}^{*} \cdot \nabla \times \overline{\mathbf{H}} = \nabla \cdot (\overline{\mathbf{E}}^{*} \times \overline{\mathbf{H}} )$ to get
\begin{equation}
    \label{eq:prior_conj}
    \begin{split}
        \nabla \cdot (\overline{\mathbf{E}}^{*} \times \overline{\mathbf{H}} ) =& -i\omega\mu^* |\overline{\mathbf{H}}|^2 + i\omega \epsilon |\overline{\mathbf{E}}|^2 +\\& i \omega\epsilon_0\frac{\beta^2}{\omega(\omega + i\gamma)} \overline{\mathbf{E}}^{*} \cdot \nabla(\nabla \cdot \overline{\mathbf{E}}).
    \end{split}
\end{equation}
Take the complex conjugate of (\ref{eq:prior_conj}) and add the two expressions, while introducing $\mu=\mu'+ i\mu''$ and $\epsilon=\epsilon'+ i\epsilon''$ \cite{olmon} ($\epsilon',\epsilon'',\mu',\mu''$, and $\beta$ are strictly real)
\begin{equation}
    \label{eq:conj}
    \begin{split}
        \nabla \cdot (\overline{\mathbf{E}}^{*} \times \overline{\mathbf{H}} + \overline{\mathbf{E}} \times \overline{\mathbf{H}}^{*} ) = -2\omega\mu'' |\overline{\mathbf{H}}|^2 - 2\omega \epsilon'' |\overline{\mathbf{E}}|^2 +
        \\ i \omega\epsilon_0\frac{\beta^2}{\omega(\omega + i\gamma)} \overline{\mathbf{E}}^{*} \cdot \nabla(\nabla \cdot \overline{\mathbf{E}}) - i \omega\epsilon_0\frac{\beta^2}{\omega(\omega - i\gamma)} \overline{\mathbf{E}} \cdot \nabla(\nabla \cdot \overline{\mathbf{E}}^*).
    \end{split}
\end{equation}
We rewrite the third and fourth terms of the right-hand side in a more compact format
\begin{equation}
    \label{eq:conj2}
    \begin{split}
        \nabla \cdot (\overline{\mathbf{E}}^{*} \times \overline{\mathbf{H}} + \overline{\mathbf{E}} \times \overline{\mathbf{H}}^{*} ) = -2\omega\mu'' |\overline{\mathbf{H}}|^2 - 2\omega \epsilon'' |\overline{\mathbf{E}}|^2  
        \\ -i \omega ( \epsilon - \epsilon_0 )\frac{\beta^2}{\omega_p^2} \overline{\mathbf{E}}^{*} \cdot \nabla(\nabla \cdot \overline{\mathbf{E}}) + i \omega ( \epsilon^* - \epsilon_0 ) \frac{\beta^2}{\omega_p^2} \overline{\mathbf{E}} \cdot \nabla(\nabla \cdot \overline{\mathbf{E}}^*),
    \end{split}
\end{equation}
integrate along the entire volume $V$ of the material, and use the divergence theorem for the left-hand side to get
\begin{equation}
    \label{eq:int1}
    \begin{split}
        &\oint_B \mathbf{n} \cdot (\overline{\mathbf{E}}^{*} \times \overline{\mathbf{H}} + \overline{\mathbf{E}} \times \overline{\mathbf{H}}^{*} )dS = \\& -\int_V 2\omega\mu'' |\overline{\mathbf{H}}|^2 dV - \int_V 2\omega \epsilon'' |\overline{\mathbf{E}}|^2 dV  
        \\& + \int_V -i \omega ( \epsilon - \epsilon_0 )\frac{\beta^2}{\omega_p^2} \overline{\mathbf{E}}^{*} \cdot \nabla(\nabla \cdot \overline{\mathbf{E}}) dV \\& +  \int_V i \omega ( \epsilon^* - \epsilon_0 ) \frac{\beta^2}{\omega_p^2} \overline{\mathbf{E}} \cdot \nabla(\nabla \cdot \overline{\mathbf{E}}^*) dV.
    \end{split}
\end{equation}
Then, apply Green's first identity to the last two terms to get
\begin{equation}
    \label{eq:int2}
    \begin{split}
        &\oint_B \mathbf{n} \cdot (\overline{\mathbf{E}}^{*} \times \overline{\mathbf{H}} + \overline{\mathbf{E}} \times \overline{\mathbf{H}}^{*} )dS = \\& -\int_V 2\omega\mu'' \left|\overline{\mathbf{H}}\right|^2 dV - \int_V 2\omega \epsilon'' \left|\overline{\mathbf{E}}\right|^2 dV  
        \\& + \oint_B -i \omega ( \epsilon - \epsilon_0 )\frac{\beta^2}{\omega_p^2} \left(\mathbf{n} \cdot \overline{\mathbf{E}}^{*} \right) \nabla \cdot \overline{\mathbf{E}} dS
        \\& - \int_V -i \omega ( \epsilon - \epsilon_0 )\frac{\beta^2}{\omega_p^2}  \left|\nabla \cdot \overline{\mathbf{E}}\right|^2 dV
        \\& + \oint_B i \omega ( \epsilon^* - \epsilon_0 ) \frac{\beta^2}{\omega_p^2} \left( \mathbf{n} \cdot \overline{\mathbf{E}}\right) \nabla \cdot \overline{\mathbf{E}}^* dS
        \\& - \int_V i \omega ( \epsilon^* - \epsilon_0 ) \frac{\beta^2}{\omega_p^2} \left|\nabla \cdot \overline{\mathbf{E}}\right|^2 dV.
    \end{split}
\end{equation}
We note that \emph{all} surface integrals vanish, by virtue of the homogeneous versions of (\ref{eq:bound1}),  (\ref{eq:bound3}), or (\ref{eq:bound2_alt}). Collecting the remaining (volume) integrals, we get
\begin{equation}
    \label{eq:int3}
    \begin{split}
       & \int_V 2\omega\mu'' \left|\overline{\mathbf{H}}\right|^2 dV + \int_V 2\omega \epsilon'' \left|\overline{\mathbf{E}}\right|^2 dV +
        \\& \int_V 2 \omega \epsilon'' \frac{\beta^2}{\omega_p^2}  \left|\nabla \cdot \overline{\mathbf{E}}\right|^2 dV = 0.
    \end{split}
\end{equation}
As long as $\mu'' \epsilon'' > 0$, it is clear that all three integrands are either nonnegative or nonpositive. In order for (\ref{eq:int3}) to hold then, there is no other possibility than $\overline{\mathbf{H}}=\overline{\mathbf{E}}=\mathbf{0}$ and the system (\ref{eq:faraday})--(\ref{eq:ampere}) accepts a unique solution. The extension reported in the following Section will complicate things; we construct Table \ref{tab:sole} where we include all the formal material conditions that ensure uniqueness in the standard, local electromagnetics, and in nonlocal, within the HDM (with real or complex hydrodynamic parameter).

While (\ref{eq:int3}) involves general constitutive parameters, we underline that the particular form of the transverse electric permittivity is known within the HDM, and neglecting the effects of bound electrons, follows the simple Drude model. Similarly, we assumed only formally a complex magnetic permeability; at optical frequencies the working materials are nonmagnetic. By examining the imaginary part of the Drude permittivity, it is simple then to establish uniqueness for the common materials that are studied by the HDM.

Some aspects of the derivation deserve comments. We note that, the introduction of nonlocality \emph{does not affect} in the slightest the material requirements set by the uniqueness theorem for local media. This is easily seen, since for \mbox{$\beta \rightarrow 0$} the third term (which includes only the divergence of the electric field, and thus only the longitudinal plasma waves predicted by the HDM) vanishes and the remaining terms are the ones found in standard literature \cite{kong}. At the same time, the mathematical details are only slightly more complicated than for the standard, local materials.

When the tangential components of both the magnetic and the electric field are prescribed on the whole surface $B$, books discussing the proof for local media underline that it suffices to use just one (tangential electric \emph{or} tangential magnetic) to guarantee uniqueness \cite{stratton,kong,felden}. This is clear from the left--hand side of (\ref{eq:int2}). Similarly we note that, for the additional surface integrals on the right--hand side of (\ref{eq:int2}) to vanish, we need $\mathbf{n} \cdot \overline{\mathbf{E}} = 0$ \emph{or} $\beta^2 \nabla \cdot \overline{\mathbf{E}}/\omega_p^2 = 0$. In the local case, as noted by Stratton \cite{stratton}, this is a puzzling aspect of the proof as it seems to contradict common practice (where both the tangential components of the magnetic \emph{and} electric field are required to deduce a solution). Taking into account the debate on the ABCs (particular form and number) the situation becomes even more confusing for nonlocal media. We clarify that the derivation presented here, \emph{by accepting a priori the boundary conditions} instead of extracting them from the particular form of the surface integrals (in order for them to vanish), does not enter in such discussions. Besides, the apparent confusion can be overcome by the same arguments that Stratton \cite{stratton} used for the local case. The field quantities entering the boundary conditions (\ref{eq:bound1}), (\ref{eq:bound3}), and  (\ref{eq:bound2_alt}) are the ``resultant'' fields, determined by the appropriately formulated boundary--value problem, which applies the practical boundary conditions -- \mbox{$\mathbf{n} \times (\mathbf{E}_{\rm out} - \mathbf{E}_{\rm in}) =\mathbf{0}$}, \mbox{$\mathbf{n} \times (\mathbf{H}_{\rm out} - \mathbf{H}_{\rm in}) =\mathbf{0}$}, \mbox{$\mathbf{n} \cdot (\epsilon_{\rm out}\mathbf{E}_{\rm out} - \mathbf{E}_{\rm in}) = 0$}, and \mbox{$ \beta^2_{\rm out} \nabla \cdot \mathbf{E}_{\rm out}/\omega_{p, \rm out}^2 - \beta^2_{\rm in} \nabla \cdot \mathbf{E}_{\rm in}/\omega_{p, \rm in}^2 = 0$} --, connecting field distributions across a surface of discontinuity. After the total fields on each side are determined, then the uniqueness theorem states that there can be no other way, as long as certain field components are specified on the boundary.

We stated earlier that the derivation was carried out under the assumption of a nonlocal background; the result of it is encapsulated in the ABCs (\ref{eq:bound2})--(\ref{eq:bound2_alt}). For the frequent case of a nonlocal--local interface the procedure remains, of course, the same. Then, we would only list (\ref{eq:bound2}) (or (\ref{eq:bound2_alt})) as the single ABC which, as discussed in the previous paragraph, suffices to guarantee uniqueness (under the same material requirements).

\section{Extension for complex $\beta^2$}
\label{sec:ext}
The extension reported herein pertains still at simple metals, but uses a complex hydrodynamic parameter. A complex $\beta^2$ is used by the GNOR model \cite{gnor} but pertains to the simple HDM as well. In the latter case, while $\beta^2$ remains real for adequately high and low frequencies, an imaginary part emerges for intermediate one \cite{halevi}. 

We employ (modifications of) the ABCs in (\ref{eq:bound2})--(\ref{eq:bound2_alt}). As before, we repeat that for the nonlocal-nonlocal interface the selection of ABCs is not trivial nor standard; the Sauter ABC though survives the introduction of a complex $\beta^2$ in GNOR \cite{gnor}.

\begin{table*}[t!]
    \centering
    \caption{MATERIAL CONDITIONS THAT GUARANTEE UNIQUENESS PER MODEL. THE SYMBOL $\gtrless$ MEANS EITHER GREATER OR LESS THAN, HOWEVER EACH CHOICE APPLIES IN TANDEM TO THE WHOLE COLUMN.}
    \begin{tabular}{c c c c }
         \hline
                    & Local & \thead{HDM for Pure Drude Metals} &  \thead{HDM with complex $\beta^2$/GNOR}\\   
        \hline
        \multirow[c]{3}{*}{Conditions} & $\mu'' \gtrless 0$ & $\mu'' \gtrless 0$ & $\mu''\gtrless 0$ \\[5pt]
         & {$\epsilon'' \gtrless 0$} & {$\epsilon'' \gtrless 0$} & {$\epsilon'' \gtrless 0$}\\[5pt]
        & & & $(\epsilon' - \epsilon_0)b'' + \epsilon'' b' \gtrless 0 $ \\[5pt]
        \hline
    \end{tabular}
    \label{tab:sole}
\end{table*}

The derivations do not formally change if we assume that \mbox{$\beta^2 = b' +ib''$}, with $b',b''$ real numbers, up until (\ref{eq:conj})
\begin{equation}
    \label{eq:conj_bcomp}
    \begin{split}
        \nabla \cdot (\overline{\mathbf{E}}^{*} \times \overline{\mathbf{H}} + \overline{\mathbf{E}} \times \overline{\mathbf{H}}^{*} ) = -2\omega\mu'' |\overline{\mathbf{H}}|^2 - 2\omega \epsilon'' |\overline{\mathbf{E}}|^2 +
        \\ i \omega\epsilon_0\frac{\beta^2}{\omega(\omega + i\gamma)} \overline{\mathbf{E}}^{*} \cdot \nabla(\nabla \cdot \overline{\mathbf{E}}) - i \omega\epsilon_0\frac{{\beta^2}^{*}}{\omega(\omega - i\gamma)} \overline{\mathbf{E}} \cdot \nabla(\nabla \cdot \overline{\mathbf{E}}^*).
    \end{split}
\end{equation}
The next steps (volume integration, divergence theorem, Green's first identity, and finally invocation of the boundary conditions) remain the same, but the corresponding to (\ref{eq:int3}) equation is more involved, namely
\begin{equation}
    \label{eq:int3_bcomp}
    \begin{split}
       & \int_V 2\omega\mu'' \left|\overline{\mathbf{H}}\right|^2 dV + \int_V 2\omega \epsilon'' \left|\overline{\mathbf{E}}\right|^2 dV +
        \\& \int_V 2 \omega [ \epsilon'' b' + ( \epsilon' -\epsilon_0 )b'' ] \frac{1}{\omega_p^2}  \left|\nabla \cdot \overline{\mathbf{E}}\right|^2 dV  = 0.
    \end{split}
\end{equation}

For $b'' = 0$ the result of (\ref{eq:int3}) is retrieved. To satisfy uniqueness, the interplay of the hydrodynamic parameter with $\epsilon$ and $\mu$ must be carefully examined. In Table \ref{tab:sole} we collect all the necessary conditions. 

As discussed previously, the permittivity is given by the Drude model. Thus, it is simple to confirm that $\epsilon''>0$ and $\epsilon'-\epsilon_0<0$ for the entire spectrum. On the other hand, the hydrodynamic parameter, according to Halevi \cite{halevi,new_halevi}, remains on the real axis for very low ($\omega \ll \gamma$) and very high frequencies ($\omega \gg \gamma$) but moves off this axis for intermediate ones, according to
\begin{equation}
    \label{eq:halevi}
    \beta^2(\omega) = \frac{ \frac{ 3 }{ 5 }\omega + \frac{ 1 }{ 3 }i\gamma }{ \omega + i\gamma} \upsilon_F^2.
\end{equation}
It is simple to deduce that the real and the imaginary part of the expression above are strictly positive and negative respectively. Similarly, in the modern GNOR formalism the generalized hydrodynamic parameter is a complex number, given by \cite{gnor}
\begin{equation}
    \label{eq:gnor}
    \eta^2 = \beta^2 + D(\gamma - i\omega),
\end{equation}
with $D$ being the diffusion constant and $\beta^2$ remaining real. Still, the real part remains strictly positive and the imaginary part strictly negative. negative, i.e., $b'>0$ and $b''<0$, in agreement with more recent and advanced perspectives \cite{newgnor}, where the hydrodynamic parameter is associated to the Fiebelman $d$ parameters at least in the frequency range where such an approach is unambiguous. We establish thus that uniqueness is guaranteed for the materials that concern GNOR/HDM with complex hydrodynamic parameter.

We note that uniqueness (of the electric field) is still controlled by the losses (whether $\epsilon''$ approaches zero or not). However, the apparition of $b''$ in the conditions is physically intuitive: in the framework of GNOR, it is linked with additional damping mechanisms, either classical and bulk (in the original paper \cite{gnor}) or quantum mechanical and surface \cite{newgnor}. Actually, the fingerprint of GNOR is size dependent line-broadening of the optical response from metallic nanoparticles, very much like resonance (blue)shift is the fingerprint of HDM \cite{gnor}. The additional loss mechanism enters (elegantly) the material conditions for uniqueness.

\section{Numerical Experiments}
\label{sec:num}

In this Section we present two numerical experiments that support the findings of the previous (theoretical) Sections.

\begin{figure}[t!]
    \centering
    \includegraphics[width=\columnwidth]{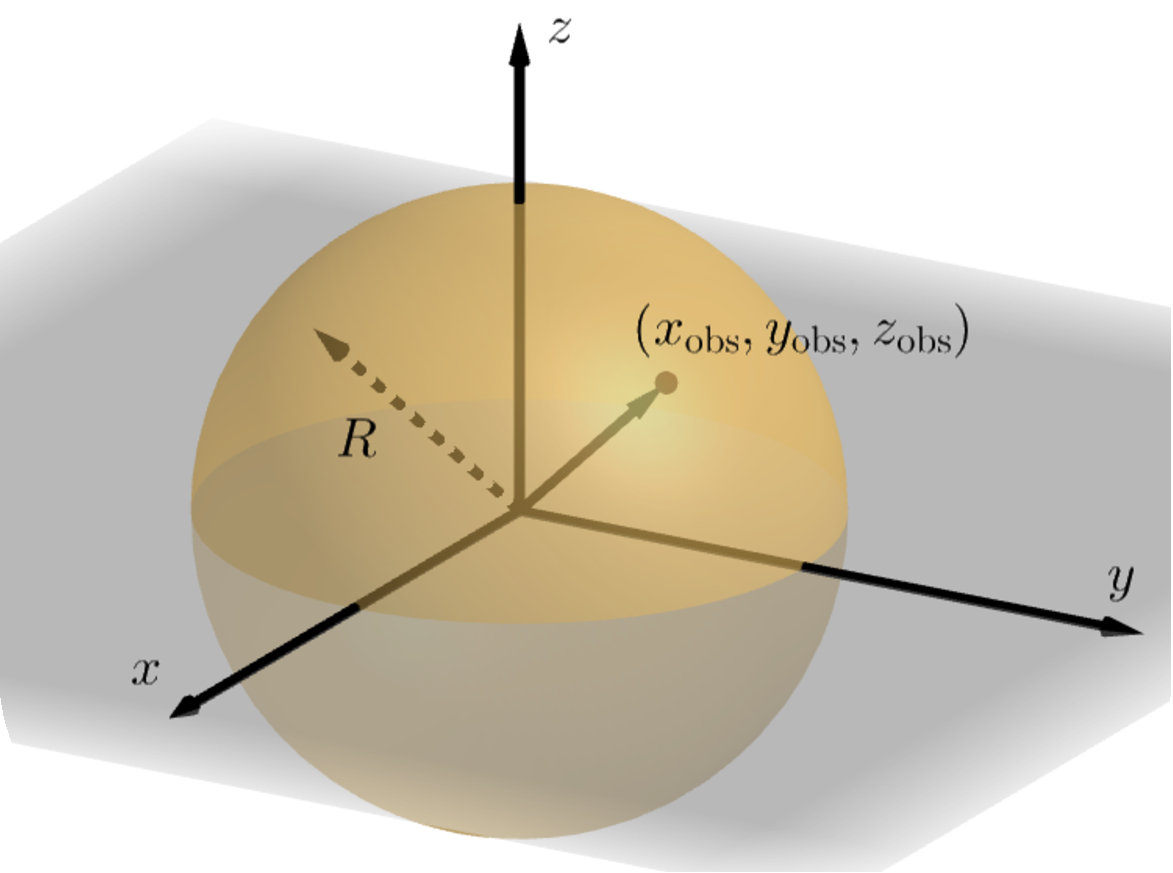}
    \caption{The nanocavity of the first numerical example---a perfect sphere with radius $R$. We adopt material parameters appropriate for gold \cite{rakic,mermin}, however we vary artificially the material losses. Fields are calculated at observation points inside the sphere described by $(x_{\rm obs},y_{\rm obs},z_{\rm obs})$.}
    \label{fig:pec}
\end{figure}

\emph{Cavities} appear as a rather fitting candidate to be tested for (non)uniquness; after all, the very problem discussed in the Sections above is an \emph{internal} and \emph{closed} one. Furthermore, a cavity that is spherical allows for further analytical evaluation. The setup (see Fig. \ref{fig:pec}) used herein is a spherical nanocavity; inside, it is filled with a nonlocal medium, modeled by the HDM and following the pure Drude model with parameters appropriate for gold \cite{rakic,mermin}, while outside it is shielded by a classical Perfect Electrical Conductor (PEC). The radius is denoted by $R$.

We admit and stress that the physicality and realism of this example is dubious. The coexistence of an (over)idealized PEC with a granular nonlocal metal seems conflicting. The motivation is to construct the simplest possible example for which the material conditions of Section \ref{sec:der} are satisfied. The interface between the nonlocal material and the PEC is more correctly a mathematical surface than a physical one, a locus where the boundary conditions discussed below hold. The second example attacks a realistic geometry.

\subsection{Spherical Nanocavity}
\label{subs:cavity}

We return to the problem at hand. A first interesting point is that we must retreat from (\ref{eq:bound2_alt}) to (\ref{eq:bound2}). Since the PEC is treated as a local medium, this is tantamount to invoking the standard Sauter ABC, namely \mbox{$\mathbf{n} \cdot \mathbf{P}_f = 0$}. For one, the use of the field ABC \mbox{$\mathbf{n} \cdot \mathbf{E}_{\rm in} = \mathbf{n} \cdot \epsilon_{\rm PEC} \mathbf{E}_{\rm PEC}$} is complicated by the infinite permittivity inside the PEC. For another, the vanishment of the current on the surface agrees with intuition: on the side of the PEC the charge collapses on the surface, and as such there is no driving force that should cause depletion inside of it.

\begin{figure*}[t!]
    \centering
    \includegraphics[width=18cm, height = 9cm]{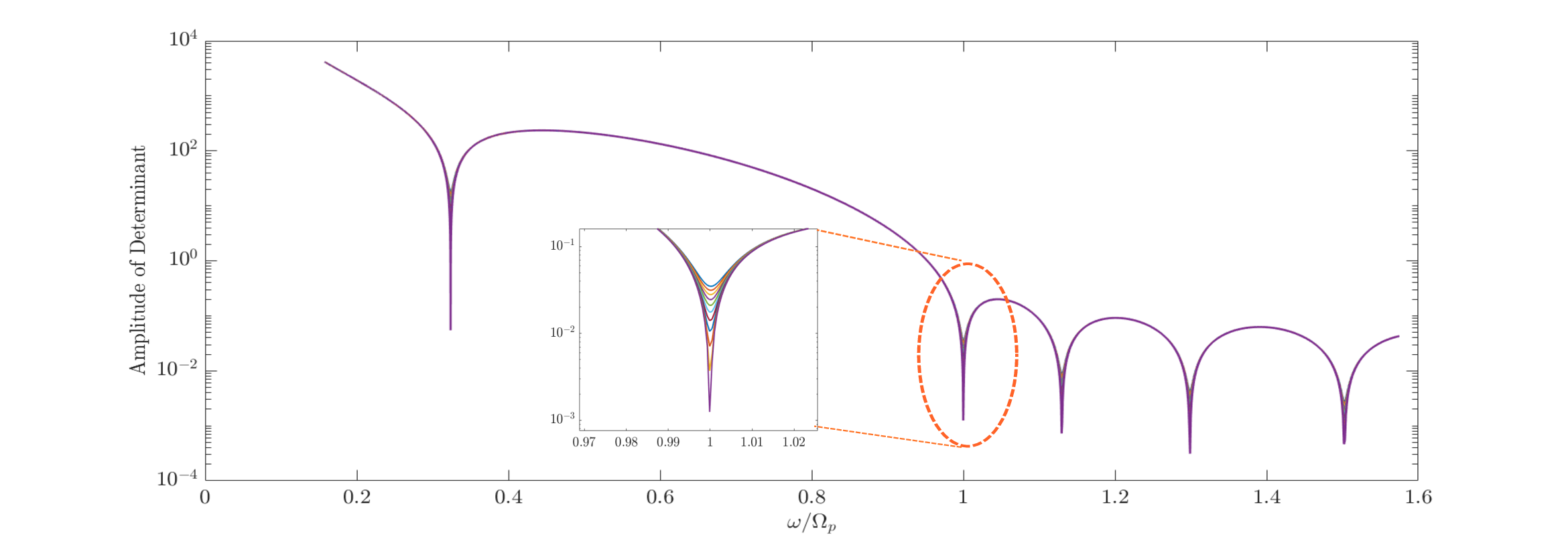}
    \caption{The determinant $\det{(\mathbf{A})}$ as a function of angular frequency, normalized to the screened plasma frequency of gold is depicted for $11$ values of the damping rate, ranging from the experimental value for gold $\gamma_{\rm exp}$ to the lossless case, when $\gamma = 0$. In the inset, a zoom-in of the second resonance is shown. The resonance becomes deeper as the damping rate decreases (the colors from up to down correspond to decreasing damping rate).}
    \label{fig:det}
\end{figure*}

The electromagnetic field inside the homogeneous and closed domain can be written as series of \emph{vector wave functions}, which are solutions to the homogeneous vector Helmholtz function \cite{stratton}, in particular
\begin{equation}
    \label{eq:series}
    \begin{split}
        &\mathbf{E}(\mathbf{r}) = \sum_{l=1}^{\infty} \sum_{m=-l}^{l} a_{lm} \mathbf{M}_{lm}(\mathbf{r}) + b_{lm} \mathbf{N}_{lm}(\mathbf{r}) + c_{lm} \mathbf{L}_{lm}(\mathbf{r}), \\
        &\mathbf{H}(\mathbf{r}) = \frac{1}{i\zeta}\left[\sum_{l=1}^{\infty} \sum_{m=-l}^{l}a_{lm} \mathbf{N}_{lm}(\mathbf{r}) + b_{lm} \mathbf{M}_{lm}(\mathbf{r}) \right],
    \end{split}    
\end{equation}
where $\mathbf{M}_{lm}$, $\mathbf{N}_{lm}$, and $\mathbf{L}_{lm}$ are the vector wave functions (defined in Appendix \ref{app:appvector}), $(l,m)$ are the the orbital quantum number and the magnetic quantum number respectively, and $\zeta = \sqrt{\mu/\epsilon}$ is the wave impedance. The series coefficients $a_{lm}$, $b_{lm}$, and $c_{lm}$ are set by the boundary conditions and the excitation.

The arbitrariness of the excitation gives us the freedom to \emph{assume} that it gives rise to a single mode $(l,m)$ with $a_{lm}=0$, that is (and focusing on the electric field)
\begin{equation}
    \label{eq:mode}
    \mathbf{E}(\mathbf{r}) = b_{lm}\mathbf{N}_{lm}(\mathbf{r}) + c_{lm} \mathbf{L}_{lm}(\mathbf{r}). 
\end{equation}
A mode that is completely described by $\mathbf{N}_{lm}$ is called \emph{transverse magnetic} (TM) \cite{stratton}. Excitations that give rise to TM modes can excite longitudinal fields as well \cite{mohan} (which are described by $\mathbf{L}_{lm}$), hence the form of (\ref{eq:mode}).

On the boundary $r=R$ of the sphere, the continuity of the tangential components of $\textbf{E}$ is supplemented by the vanishement of the normal components of $\mathbf{P}_f$, following from (\ref{eq:bound2}). The $2 \times 2$ system that ensues is (see Appendix \ref{app:matrix})
\begin{equation}
    \label{eq:det}
    \underbrace{\begin{bmatrix}
         \displaystyle \frac{ 1 }{ kr } \frac{\partial \left[ r j_l( k r)  \right]}{\partial r} &  \displaystyle \frac{ j_l( \kappa r )}{ \kappa r } \\
         \displaystyle l ( l + 1 ) \left( \frac{\epsilon}{\epsilon_0} - 1 \right) \frac{ j_l( k r )}{ k r } & - \displaystyle \frac{ \partial j_l( \kappa r )}{ \partial( \kappa r ) }
    \end{bmatrix}}_{\mathbf{A}}
    \underbrace{\begin{bmatrix}
        b_{lm} \\
        c_{lm}
    \end{bmatrix}}_{\mathbf{x}} = 
    \underbrace{\begin{bmatrix}
        0 \\
        0
    \end{bmatrix}}_{\mathbf{b}},
\end{equation}
at $r=R$. Above, $k(\omega) = \omega \sqrt{\epsilon \mu}$ is the standard transverse wavenumber, and \cite{forstmann}
\begin{equation}
    \label{eq:kappa}
    \kappa(\omega)=\frac{ 1 }{ \beta } \left[ \omega( \omega + i\gamma ) - \omega_p^2  \right]^{\frac{1}{2}},
\end{equation}
is the longitudinal wavenumber as it ensues from the dispersion relation for longitudinal waves, in particular $\epsilon(\omega,\mathbf{k})=0$ \cite{maier}. For convenience, we denote this system as $\mathbf{A} \cdot \mathbf{x} = \mathbf{0}$. The system is quite similar to the one presented in \cite{mario}; note that the nonlocal sphere in this work is embedded in a dielectric. We note in passing that, for a purely local cavity, when $c_{lm}=0$ and the ABC is neglected, the only element of $\mathbf{A}$ surviving is the upper left one ($11$), as expected (compare, after some simplifications, with \cite[Eq.~33,~p.~560]{stratton} for the ``electric modes'').

Aside from the trivial solution $(b_{lm},c_{lm})=(0,0)$, additional ones may arise when the system matrix $\mathbf{A}$ is nonivertible, that is, $\det{(\mathbf{A})} = 0$. The frequencies corresponding to solutions of this equation are the \emph{resonant frequencies}, complex in general $\omega_r = \omega'+i\omega''$. The imaginary part arises from material and radiative losses; the use of the (rather unrealistic) PEC aims at suppressing the latter and allows us to focus on the former, which we discuss in our derivations above. From the discussion in Section \ref{sec:der}, and since $\mu = \mu_0$, we expect that uniqueness breaks down only if $\epsilon'' \rightarrow 0$. In turn, this is achieved if $\gamma=0$.

In Fig. \ref{fig:det} we plot the frequency-dependent determinant $\det{(\mathbf{A})}$ along \emph{real} frequencies for $11$ different damping rates, for $10,000$ frequency points between $100$ nm and $1,000$ nm, and for $l=1$ (the dipolar mode will be the one studied in all subsequent experiments). We employ a simple formula to progressively diminish the losses to zero, in particular
\begin{equation}
    \label{eq:gamma}
    \gamma_j = \gamma_{\rm exp} - \frac{j}{N} \gamma_{\rm exp}, \quad j=0,1,\dots, N,
\end{equation}
where $N+1$ is the number of different damping rates studied (for this scenario $N=10$) and $ \hbar \gamma_{\rm exp} = 0.053$ eV ($\hbar$ being the reduced Planck constant) is a realistic starting value corresponding to gold~\cite{rakic}. We study a tiny sphere of $R=1$ nm; we observe the same behavior for larger radii, however for demonstration purposes we select a small one, that further enhances the observations we want to make. The remaining essential parameters are the screened plasma frequency \mbox{$\hbar \Omega_p = \hbar\sqrt{0.760}\omega_p = 7.87$} eV \cite{rakic} and the Fermi velocity $\upsilon_F = 1.40 \times 10^6$ m/s \cite{mermin}, suitable for gold.

Two aspects of Fig. \ref{fig:det} deserve to be commented. For one, all cases exhibit acute dips. The frequencies on which these dips happen correspond to the real part of the resonant frequencies $\omega'$ (note that they appear normalized to the screened plasma frequency. For another, the dips are deeper and cruder as $\gamma \rightarrow 0$ (see the inset in Fig. \ref{fig:det}). Therefore in the driven case, the coefficients $(b_{lm},c_{lm})$ and accordingly the electromagnetic field will be determined through $\mathbf{A}^{-1}$. Clearly, when $|\det{(\mathbf{A})}| \rightarrow 0$ they will exhibit higher and higher and eventually undefined amplitudes on the resonant frequencies; the zeroes of the determinant translate to singularities for the field. 

\begin{figure}[t!]
    \centering
    \includegraphics[width=\columnwidth]{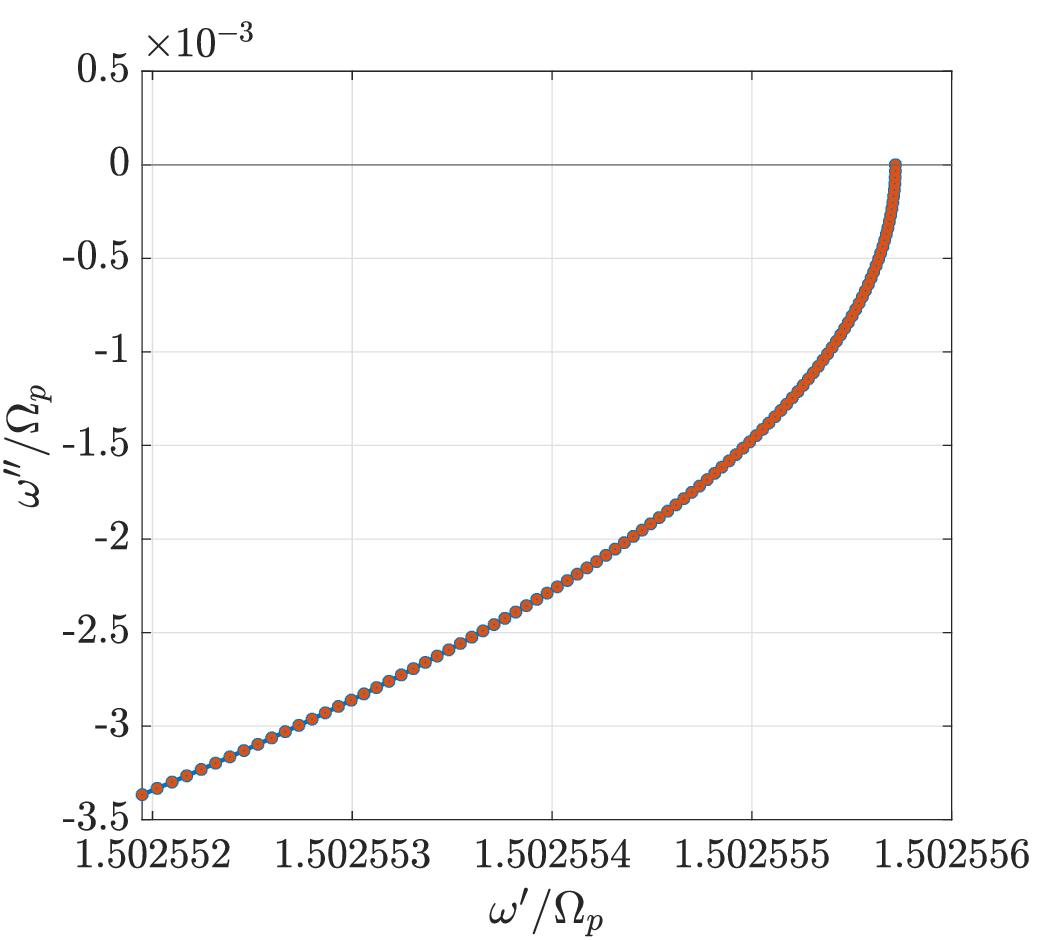}
    \caption{The imaginary part $\omega''$ versus the real part $\omega'$ of the resonant frequency for the largest (fourth) resonance of Fig. \ref{fig:det}. The blue line is a guide to the eye; the blue points correspond to the (complex) resonant frequency for various losses, as described by (\ref{eq:gamma}) for $N=100$ as calculated by the minimizer. The orange circles correspond to the imaginary part being calculated by $-\gamma/2$. The thin black line underlines the real axis.}
    \label{fig:grid}
\end{figure}

Using the real part of the resonant frequencies collected from Fig. \ref{fig:det} as input to a minimizer (indicatively for the largest resonant frequency) and guessing an imaginary part of $-10$ THz\footnote{This choice, corresponding to $\approx -0.001 \Omega_p$, allows for values of $|\det{(\mathbf{A})}|$ in the order of $10^{-15}-10^{-16}$ depending on $\gamma$.} we can extract the complete (within some tolerance) complex resonant frequency that nullifies $|\det{(\mathbf{A})}|$. As shown in the curve in blue in Fig. \ref{fig:grid}  (for $100,000$ frequency points and $101$ loss points) the resonant frequencies are spilled in the complex plane for all lossy ($\gamma \neq 0$) cases, but almost (within a few Hz\footnote{We speculate that this (minimal) discrepancy rises from the selection of the starting point. If the starting imaginary part is selected large as described in footnote 1, the resonant frequency for $\gamma=0$ ends up with a small positive imaginary part equal to $\approx 1.15$ Hz. For a purely real starting point, the imaginary part is squeezed to a mere $1.27 \times 10^{-4}$ Hz.}) fall on the real axis for the lossless case.

A comparison between our cavity and standard system theory \cite{chew_notes} reveals why the behavior of Fig. \ref{fig:grid} is an indication of non-uniqueness in the lossless case. Rewrite the system of (\ref{eq:det}) as
\begin{equation}
    \label{eq:sys}
    \mathbf{x}(\omega) = \left( \frac{\rm{adj}{(\mathbf{A})}}{\det{(\mathbf{A})}} \right) (\omega) \cdot \mathbf{b}(\omega),
\end{equation}
where $\mathbf{b}$ describes the excitation ($\mathbf{0}$ for the non-driven case) and $\rm{adj}$ denotes the adjunct matrix. The equation above resembles much the standard input-output equation for linear systems $Y(\omega)=H(\omega)X(\omega)$ where $Y$ is the output, $X$ the input, and $H$ the transfer function, which describes the behavior of the system, just as the matrix determinant does in our case. Taking the cue from \cite{chew_notes}, we realize that the Fourier transform $(\rm{adj}(\mathbf{A})/\det{(\mathbf{A}))(\omega)}$ is not defined, should a pole be encountered on the integration path. This is exactly the case when $\gamma = 0$ and this is the origin of forcing \emph{indented} paths of integration even in local computational electromagnetics \cite{chew}, to ensure uniqueness and thus improve numerical stability.

We notice as well that the ``numerical'' predictions of the imaginary part of the resonant frequencies, agree well with a theoretical result (see orange circles, Fig. \ref{fig:grid}), namely $\omega''=-\gamma/2$, typically discussed for open geometries \cite{raza}, though recently rediscovered for the nonlocal cavity in \cite{filipa}. The maximum relative error between the imaginary part of the output of the minimizer and the theoretical prediction is a mere $6.54 \times 10^{-5} $ $\%$ (exempting the lossless case, when the numerical prediction \emph{almost} falls on the real axis, see footnote 2).
\begin{figure}[t!]
    \centering
    \includegraphics[scale=0.51]{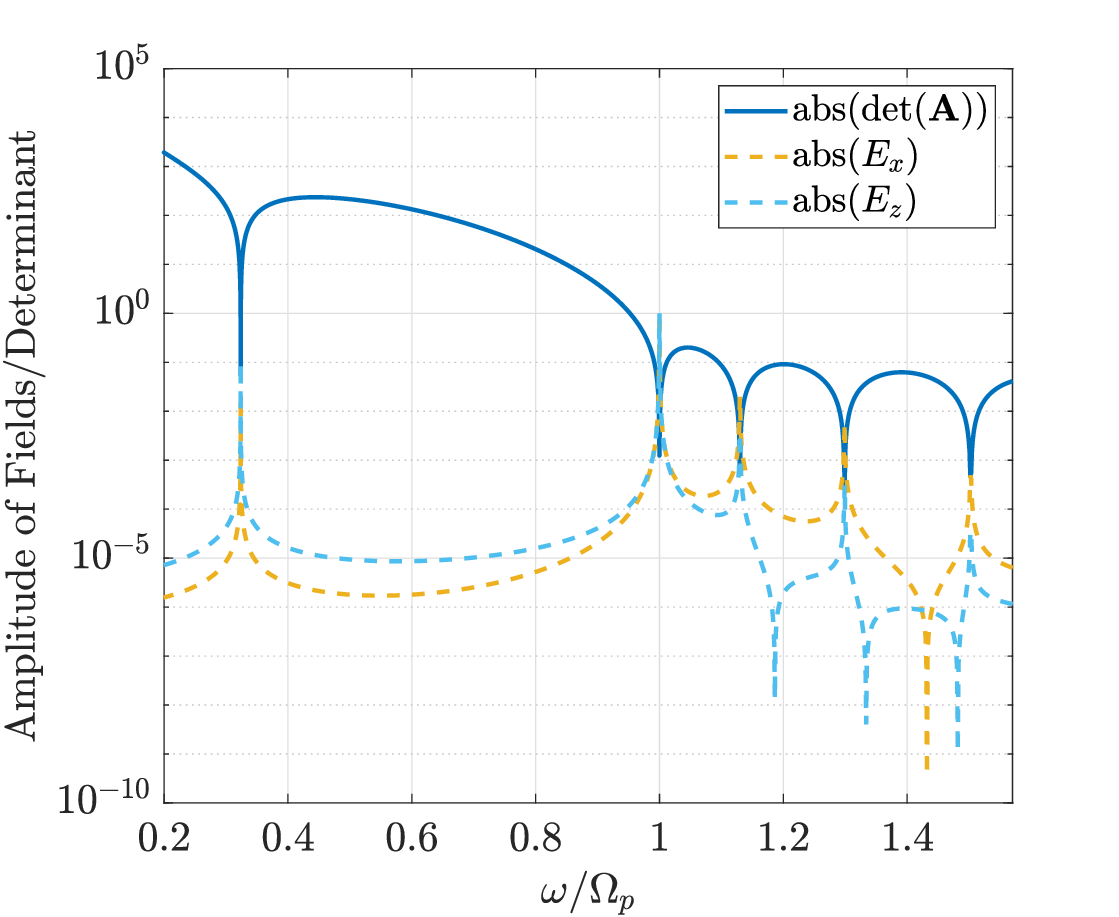}
    \caption{The absolute value of the determinant in the same graph with the normalized (to their maximum value) components of the field distribution. The field is calculated at $(x,y,z)=(0.5,0,0.1)$ nm, which is a point inside the cavity. The thick blue line corresponds to the absolute value of the determinant, as in Fig. \ref{fig:det}. The dashed lines denote the field components and in particular, $x$ with yellow and $z$ with cyan. All results are gathered for $\gamma=0$.}
    \label{fig:field_det}
\end{figure}
All the above are indications that uniqueness breaks down in the lossless case, as expected by the theoretical treatment of Section \ref{sec:der}. Perhaps, the clearest manner to confirm it is to study the electromagnetic field at a point of the cavity for the driven case. As an excitation, we select a spherical wave that drives TM and longitudinal modes, in particular
\begin{equation}
    \label{eq:exc}
    \mathbf{E}_{\rm inc}(\mathbf{r}) = \mathbf{N}_{10}(\mathbf{r}).
\end{equation}
See Appendix \ref{app:appvector} for more information on $\mathbf{N}$ and note that the spherical Hankel function of the first kind is selected for it in order to describe a spherical wave generated by a singularity at $r=0$ and impinging on the inner side of the spherical interface. The vector of the excitation then becomes (see Appendix \ref{app:matrix})
\begin{equation}
    \label{eq:bexc}
    \mathbf{b} = 
    \begin{bmatrix}
        \displaystyle -\frac{1}{kr} \frac{ \partial \left[ r h_1^{(1)}( k r )\right]}{ \partial r }  \\
        
        \displaystyle -2 \left( \frac{\epsilon}{\epsilon_0} - 1 \right) \frac{h_1^{(1)}( k r )}{k r } 
    \end{bmatrix}, \quad r=R.
\end{equation}
Using the orthonormality properties of vector and scalar spherical harmonics (see Appendix \ref{app:appvector}) we deduce that a single mode, $(l,m)=(1,0)$ is excited. With this choice of $(l,m)$, we combine (\ref{eq:bexc}) with $\mathbf{A}$ of (\ref{eq:det}) and solve for $(b_{10},c_{10})$. The scattered electric field is then determined at any point $\mathbf{r}$ inside the cavity via (\ref{eq:mode}). In Fig. \ref{fig:field_det} we plot, for the lossless case, the determinant of the system's matrix, as well as the $x$ and $z$ components of the field distribution, normalized to their maximum value (to fit the scale). As for Fig. \ref{fig:det}, $10,000$ wavelengths are sampled and $l=1$.
Note that $E_y=0$, as expected from the excitation for this observation point (where $\phi=0$).

\begin{figure}[t!]
    \centering
    \includegraphics[scale=0.55]{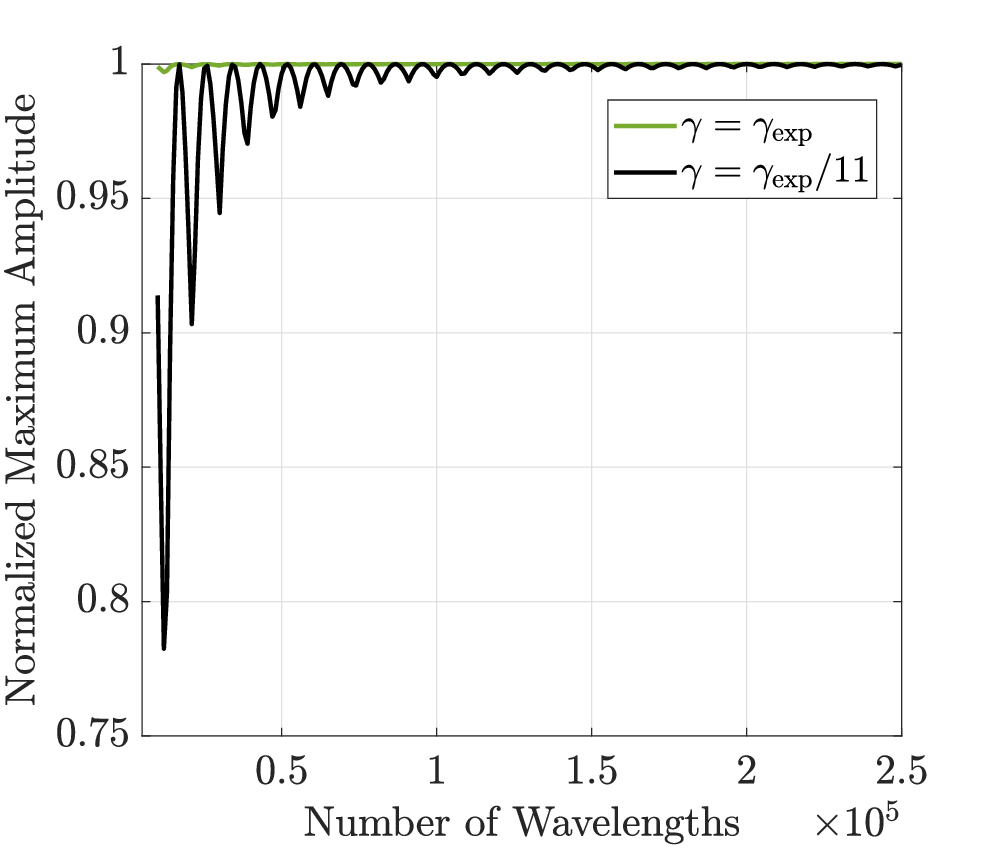}
    \caption{The maximum amplitude for $|E_z|$ is plotted against the number of frequency points (the refinement of the discretizetation) for $\gamma=\gamma_{\exp}$ (green dots) and for $\gamma=\gamma_{\exp}/11$ (black line). }
    \label{fig:conv}
\end{figure}

It becomes immediately obvious that at the resonant frequencies the field demonstrates local maxima (and quite powerful ones when the normalization is removed), as expected from (\ref{eq:sys}). The remaining features of the curves (i.e., the sharp local minima) are not of concern to the arguments we are using here. Intuitively, we anticipate that this rough landscape is created by the multiple reflection of the scattered wave on the conducting walls of the cavity and the singularity introduced at $r=0$ by the excitation.

Since the fields exhibit maxima at the ``pathological'' resonant frequencies, it is a natural step to probe, by increasing the discretization (namely the number of sampled wavelengths), how a particular field maximum can be reconstructed. In Fig. \ref{fig:conv} we showcase the convergence to the maximal value for two different values of $\gamma$, and for $|E_z|$. We sweep between different discretizations, ranging from $10,000$ to $250,000$ frequency points with a step of $1,000$ and for the initial interval of $100$ to $1,000$ nm. We have confirmed that $|E_x|$ demonstrates the same behavior. Raising the number of sampled frequencies, we expect to progressively fully reconstruct the behavior of the field on its dominant resonance. This is achieved rather easily for the realistic case (with $\gamma_{\rm exp}$). Much more challenging and oscillatory in nature is the convergence of the case with the diminished losses. The convergence is clear, however a very fine discretization is required. In both cases, no matter how slow the convergence is, uniqueness is achieved. Since the green, lossy curve may give an impression of flat-lining, we would like to clarify that both curves showcase such oscillatory behavior, with the lossy one being much milder, and thus being considered as converging faster. For the case when $\gamma = 0$, the results appear noisy, with no sense of convergence. In other words, the value of the field at this point is undefined and consecutive simulations may give similar or radically different results. We consider this the most definite indication that when the conditions of Table \ref{tab:sole} are violated, then the solution to Maxwell's equation becomes indeed nonunique. We close this example by providing the full code in Appendix \ref{app:code}.

\subsection{Nanocube in Vacuum}
\label{subs:cube_v}
\begin{figure}
    \centering
    \includegraphics[width=\columnwidth]{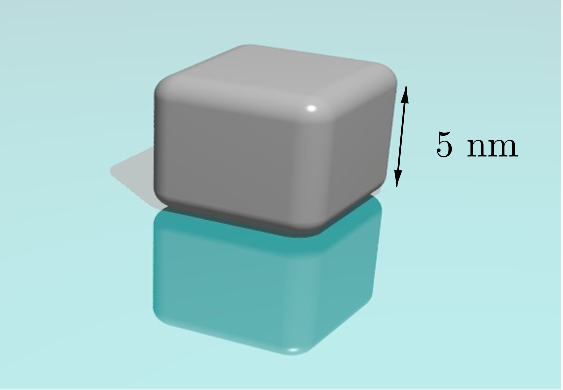}
    \caption{The solitary nanocube of the second example. Each edge has a length of $5$ nm. For the material properties, we borrow from silver and \cite{rakic,mermin,diff}. The background is considered to be vacuum (note that the plane below is just a graphical representation).}
    \label{fig:cube}
\end{figure}

We proceed with the second numerical example. The setup, shown in Fig. \ref{fig:cube}, consists of a nanocube with rounded angles. Each edge has a length of $5$ nm, a relatively small size to limit the radiated power. The structure is embedded in vacuum and is described by parameters of silver \cite{rakic}, disregarding the effects of bound electrons (i.e., adopting the pure Drude model), with the losses being modulated by (\ref{eq:gamma}). These are  $\hbar \Omega_p =  \hbar\sqrt{0.845}\omega_p = 8.28$ eV, $\hbar\gamma_{\rm exp} = 0.048$ eV and $\upsilon_F = 1.39 \times 10^6$ m/s \cite{mermin}. Additionally, we model the free-electron response by means of GNOR, for which we require the diffusion constant $D = 2.684 \times 10^{-4}$ m\textsuperscript{2}s\textsuperscript{-1} from \cite{diff}. Thus, the hydrodynamic parameters are frequency dependent and given by (\ref{eq:gnor}).
We probe the response of this nanoparticle by means of an in-house developed 3D BEM algorithm \cite{bem1,bem3}, which revolves around an electromagnetic potential-based reformulation of the governing equations for efficiency. We aim to support here our initial claim: even algorithms that only \emph{implicitly} step on Maxwell's equations and the hydrodynamic equation of motion are still enabled by the uniqueness theorem and are dictated by its terms. We will not elaborate on the details of the BEM here; they can be found in a series of papers by our group, see \cite{bem1,bem2,bem2d,bem3,bem4}. From a bird eye's view, the governing differential equations (wave equations for potentials) are transformed into integral equations and by a limiting process, to \emph{boundary} integral equations, which satisfy the required boundary conditions. Potentials and fields are generated not by means of the material contrasts of the scatterers under study, but rather by sets of equivalent currents, charges, and ``longitudinal'' charges, the latter being responsible for the new, longitudinal degree of freedom predicted by the nonlocal dynamics. These equivalent sources are the unknowns of the integral equations and are to be determined. To achieve this, the integral equations are transformed to a system of matrix equations by means of a standard MoM.

As such, in this case, quite different from the analytic toy problem discussed previously, we can create a matrix equation similar to (\ref{eq:det}), in particular \cite{bem1}
\begin{equation}
    \label{eq:bem_system}
    \underbrace{\begin{bmatrix}
        M_{11} & M_{12} \\
        M_{21} & M_{22}
    \end{bmatrix}}_{\mathbf{A}_{\rm BEM}}
    \underbrace{\begin{bmatrix}
        \sigma_{1} \\
        \sigma^L_{1}
    \end{bmatrix}}_{\mathbf{x}_{ \rm BEM}} = 
    \underbrace{\begin{bmatrix}
        b_{11} \\
        b_{21}
    \end{bmatrix}}_{\mathbf{b}_{\rm BEM}}.
\end{equation}
Above $\sigma_1$ and $\sigma_1^L$ are the equivalent charge and longitudinal charge lying on the inner side of an interface. The currents of either sides and the charges on the outer are extracted from $\sigma_1$ and $\sigma_1^L$. The matrix elements $M_{\rho\nu}$, with $\rho, \nu = 1,2$ are defined by means of complicated operator expressions; what is important to notice is their dimensions: $T \times T$ where $T$ is the number of triangular patches used to discretize the nanoparticle. The system matrix $\mathbf{A}_{\rm BEM}$ is a block matrix here. Finally, the excitation is dictated by $\mathbf{b}_{\rm BEM}$ and is zero for the nondriven case.
\begin{figure*}[t!]
    \centering
    \hspace*{-1.3cm}   
    \includegraphics[scale=0.67]{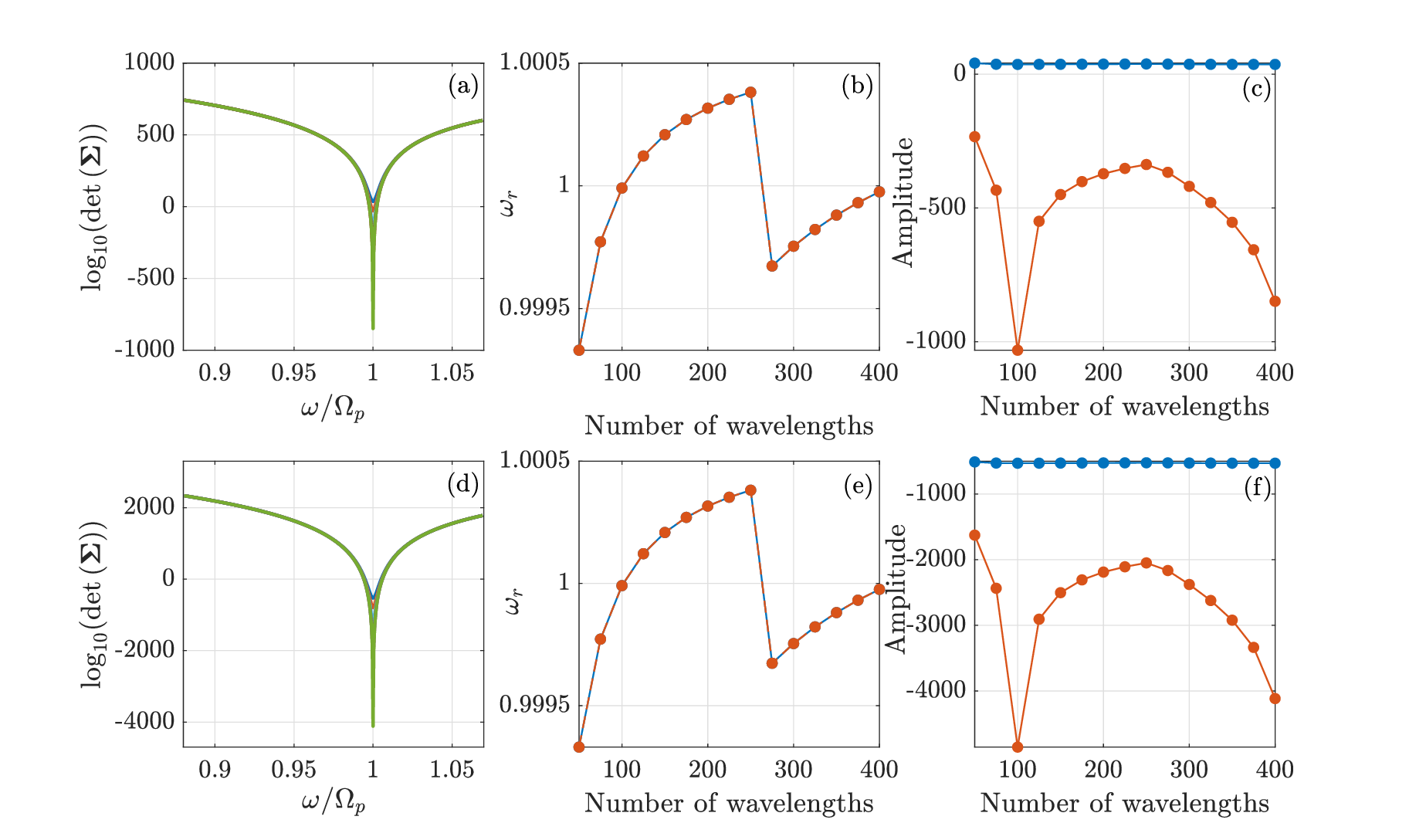}
    \caption{Numerical results based on the determinant of the system matrix in \ref{eq:bem_system}. Two different meshes, Mesh 1 (first row, sparse) and Mesh 2 (second row, dense) are used. The determinant (after being manipulated as described in this Subsection) is shown for $400$ wavelengths and $5$ different loss values, namely $\gamma=\gamma_{\rm exp}$, $\gamma=0.7\gamma_{\rm exp}$, $\gamma=0.4\gamma_{\rm exp}$, $\gamma=0.1\gamma_{\rm exp}$, and $\gamma=0$. The deeper the resonance, the smaller the loss. In (b) and (e) the location (frequency) of the resonance is depicted and is to be analyzed in conjunction with (c) and (f), where the amplitude of the resonance is shown for the lossy case $\gamma=\gamma_{\rm exp}$ in blue and for the lossless case in orange. Note that frequencies are depicted normalized to the screened plasma frequency.}
    \label{fig:numerics}
\end{figure*}
As we have discussed for the analytic example, the determinant of $\mathbf{A}_{\rm BEM}$ controls the response of the system. The computation thereof is a rather challenging task as direct evaluation (per frequency/wavelength point of the square $2T \times 2T$ matrix) results in a numerical explosion. We traced the problem to the fine mesh used. Actually, all but extremely sparse meshes could not be probed. To circumvent the problem, we performed \emph{singular value decomposition} on $\mathbf{A}_{\rm BEM}$, an approach inspired by \cite{svd1,svd2}; this leads to the diagonalization $\mathbf{A}_{\rm BEM} = \mathbf{U} \mathbf{\Sigma} \mathbf{V}^{\rm H}$ where $\rm H$, denotes the conjugate transpose and which is efficiently implemented as a built-in function in MATLAB \cite{svd}. All matrices share the dimensions of $\mathbf{A}_{\rm BEM}$ Here, $\mathbf{U}$ and $\mathbf{V}^{\rm H}$ are \emph{unitary} matrices, so $\det(\mathbf{U})=\det(\mathbf{V}^{\rm H}) = 1$. Matrix $\mathbf{\Sigma}$ is a diagonal matrix that contains the \emph{singular values} of $\mathbf{A}_{\rm BEM}$, which are chosen to be nonnegative numbers (something that we have confirmed numerically). We note then that
\begin{equation}
    \label{eq:svd_first}
    \det{(\mathbf{A}_{\rm BEM})} = \det{(\mathbf{\Sigma})} = \prod_{i=1}^{2T}{\sigma_{ii}},
\end{equation}
where $\sigma_{ii}$ are the aforesaid singular values and are not to be confused with the equivalent charges. This is a fruitless effort as well, however we can still probe the determinant, at least qualitatively, should we study the logarithm of it instead
\begin{equation}
    \label{eq:svd_second}
    \log_{10}(\det{(\mathbf{\Sigma})}) = \sum_{i=1}^{2T}{\log_{10}(\sigma_{ii})} = \rm{Tr}\left( \log_{10}(\mathbf{\Sigma}) \right),
\end{equation}
where $\rm Tr$ denotes the trace of the diagonal matrix. In other words, we are studying $\log_{10}(\det(\mathbf{A}_{\rm BEM}))$ through a numerically amenable route, since we can extract finite valued results (dependent on the mesh density). The new, fully real, determinant is expected to yield even negative values; this is not surprising as the amplitude of this determinant translates to the exponent with basis $10$ of the actual one (thus negative values should correspond to regions around the resonance(s)). 

Having set the necessary formulation, we proceed with the numerical experiment. We visualize the determinant by performing a wavelength (frequency) sweep between 140 and 170 nm. We note that the complexity of the algorithm does not allow for (almost) arbitrarily   fine discretization as for the analytic case, thus, to increase the density of the discretization, we investigate a narrower spectrum. Two different mesh densities are evaluated: the first consists of 428 triangular patches (Mesh 1), with a granular discretization of the rounded edges and corners and a sparser elsewhere, and a second with 1728 triangular patches (Mesh 2) and a more balanced distribution of the still adaptive mesh that prioritizes corners and rounded edges.

In Fig. \ref{fig:numerics} (a) and (d) we draw the said determinant for 400 frequency points for Mesh 1 and Mesh 2 respectively and we observe--as for the analytic case--that with decreasing $\gamma$ (using (\ref{eq:gamma})) the resonance deepens, even when $D \neq 0$. We note that the particular behavior of the determinant is not a caprice of the sparser Mesh 1, but is recreated completely in Mesh 2, though as mentioned before, the result is mesh dependent. This agrees well with intuition, especially in the region of the resonance, where the (far) more negative values of the determinant of Mesh 2 signify, of course, better convergence.   

As in the semianalytical case, deeper resonances with decreasing the damping rate are associated with nonuniqueness. For this purpose, we further test the behavior of the determinant on the resonant frequency, by modifying the frequency discretization. The spectrum of $130$ to $170$ nm is swept using $50$ to $400$ frequency points, with a step of $50$. In Fig. \ref{fig:numerics} (b) and (e) the frequency where the minimum is located is depicted for the case when $\gamma=\gamma_{\rm exp}$ (blue) and for the lossless case (orange). This result seems unaffected by the refinement of the mesh, but dependent on the frequency resolution. Most importantly it can be read together with Fig. \ref{fig:numerics} (c) and (f), where using the same color code, the amplitude of the determinant at its minimum is shown. The amplitude follows in both (lossy and lossless) cases the frequency trend with respect to the plasma frequency, location of the material resonance when $\gamma=0$ and $\omega=\Omega_p$. We note however, as in the first example, that the situation between the lossy and the lossless cases differs significantly in the variation of reported values: for the lossy case we calculate a standard mean deviation of $1.20$ ($4.89$ for Mesh 2) versus one of $209$ ($844.72$) for the lossy case. As in Fig. \ref{fig:conv}, both curves showcase a variation, but for the lossy case this seems to be much wilder, and the convergence (if any) slower. We cannot be as bold in our assertions here, since the purely numerical case suffers from additional complications, either physical, as the open problem will have some radiation loss, even negligible, and most importantly, computational, as the rather demanding algorithm does not allow for an as meticulous investigation as for the analytics (especially with respect to increasing the wavelength resolution). In any case, we evaluate the presented results as congruent to the intuition we built in this work, the analytical results, and the theory presented in the previous Sections; when the material becomes lossless, even if a complex hydrodynamic parameter is used, the uniqueness of the solution is jeopardized.

\section{Conclusion}

The main question answered in this work is how does the introduction of nonlocal mechanisms through hydrodynamic models, such as the HDM and GNOR, influence the material-response requirements for uniqueness of solutions to Maxwell's equations. The procedure followed is an adaptation to the complicated constitutive relations of the derivations showcased in popular electromagnetics textbooks. As such, the points of departure from local approximations are made clear. This framework is particularly effective at the same time in elucidating the role and necessity in using ABCs, in order to eliminate the surface integrals arising in the derivations, see (\ref{eq:int2}). Interestingly enough, HDM and local theory share the same material requirements, while additional ones arise with the introduction of a complex hydrodynamic parameter. Two examples, a semianalytical and a numerical one, support the thesis, the first for the HDM and the second for GNOR. The second example is especially illustrating of how the uniqueness theorem constitutes definitely if covertly a stepping stone of common practice recipes that may be significantly reformulated with respect to the original system of Maxwell's equations and the hydrodynamic equation of motion.

\label{sec:con}

\appendices

\section{Vector Wave Functions}
\label{app:appvector}

The vector wave functions are generated by the \emph{scalar} wave function, which is a solution to the homogeneous and scalar Helmholtz equation \cite{stratton}
\begin{equation}
    \label{eq:helm}
    \nabla^2 \psi_{lm}(\mathbf{r}) + K^2\psi_{lm}(\mathbf{r}) = 0,
\end{equation}
where $K$ is either the transverse or the longitudinal wavenumber depending on whether $\psi_{lm}$ describes transverse or longitudinal waves. Here we use the indices $(l,m)$ for the spherical waves; one could use the generic index $n$ instead. 

The vector wave functions are then defined as \cite{chew}
\begin{equation}
    \label{eq:M}
    \mathbf{M}_{lm}(\mathbf{r}) = \nabla \times \mathbf{r} \psi_{lm}(\mathbf{r}), 
\end{equation}
\begin{equation}
    \label{eq:N}
    \mathbf{N}_{lm}(\mathbf{r}) = \frac{1}{k} \nabla \times \nabla \times \mathbf{r}\psi_{lm}( \mathbf{r} ),   
\end{equation}
\begin{equation}
    \label{eq:L}
    \mathbf{L}_{lm}(\mathbf{r}) = \frac{1}{\kappa} \nabla \psi_{lm}(\mathbf{r}).
\end{equation}
Note that the factor $\kappa^{-1}$ in (\ref{eq:L}) is often excluded from consideration \cite{stratton}.  Subject to (\ref{eq:helm}), $\mathbf{M}_{lm}$, $\mathbf{N}_{lm}$, and $\mathbf{L}_{lm}$ are solutions to the homogeneous vector Helmholtz equation, as we stated earlier. It is further interesting to notice that $\mathbf{M}_{lm}$ and $\mathbf{N}_{lm}$ are solenoidal, thus suited for the description of transverse waves, while $\mathbf{L}_{lm}$ is irrotational, thus suited for the description of longitudinal waves.

Equation (\ref{eq:helm}) can be analytically solved, yielding \cite{chew}
\begin{equation}
    \label{eq:psi}
    \psi_{lm}(\mathbf{r}) = z_{l}(Kr)Y_{lm}(\theta,\phi),
\end{equation}
where $(r,\theta,\phi)$ are the standard spherical coordinates, $z_l$ denotes either the spherical Bessel or Hankel functions of the first kind and $Y_{lm}$ the scalar spherical harmonics.

It is straightforward albeit laborious to deduce the vector wave functions from (\ref{eq:M})--(\ref{eq:L}), leading to
\begin{equation}
    \label{eq:M2}
    \mathbf{M}_{lm}(\mathbf{r}) = z_l( k r) \mathbf{X}_{lm}(\theta,\phi) , 
\end{equation}
\begin{equation}
    \label{eq:N2}
    \mathbf{N}_{lm}(\mathbf{r}) = l ( l + 1 ) \frac{z_l ( k r) }{k r} Y_{lm}( \theta, \phi )\hat{\mathbf{r}} + \frac{1}{kr} \frac{ \partial \left[ r z_l( k r )\right]}{ \partial r } \mathbf{Z}_{lm}( \theta, \phi ), 
\end{equation}
\begin{equation}
    \label{eq:L2}
    \mathbf{L}_{lm}(\mathbf{r}) = \frac{ \partial z_l( \kappa r )}{ \partial (\kappa r )}Y_{lm}(\theta,\phi) \hat{\mathbf{r}} + \frac{ z_l( \kappa r )}{ \kappa r } \mathbf{Z}_{lm}(\theta,\phi). 
\end{equation}
$\mathbf{X}_{lm}$ and $\mathbf{Z}_{lm}$ are the (two out of the three) vector spherical harmonics. The scalar spherical harmonics are defined as \cite{nist}
\begin{equation}
    \label{eq:Y}
    Y_{lm}(\theta,\phi) = \sqrt{\frac{2l + 1}{4\pi} \frac{(l-m)!}{(l+m)!}} P_l^m(\cos{\theta})e^{ i m \phi},
\end{equation}
where ``$!$'' denotes the factorial and $P_l^m$ is the associated Legendre polynomial of degree $l$ and order $m$. On the other hand, the vector spherical harmonics are \cite{morse}
\begin{equation}
    \label{eq:vX}
    \begin{split}
        \mathbf{X}_{lm}(\theta,\phi) =& \nabla \times \left( \mathbf{r} Y_{lm} ( \theta, \phi )    \right) 
        \\=& \frac{1}{\sin{\theta}} \frac{ \partial Y_{lm}( \theta, \phi )}{ \partial \phi } \hat{\pmb{\theta}} - \frac{\partial Y_{lm}( \theta, \phi )}{\partial \theta} \hat{\pmb{\phi}}
    \end{split}
\end{equation}
and
\begin{equation}
    \label{eq:vZ}
    \begin{split}
        \mathbf{Z}_{lm}(\theta,\phi) =& r\nabla Y_{lm}( \theta, \phi ) \\
        =& \frac{ \partial Y_{lm}( \theta, \phi )}{ \partial \theta } \hat{\pmb{\theta}} + \frac{1}{\sin{\theta}} \frac{\partial Y_{lm}( \theta, \phi )}{\partial \phi} \hat{\pmb{\phi}}.
    \end{split}
\end{equation}

From the many properties of scalar and vector spherical harmonics, we include here orthonormality relations that will serve us later \cite{nist}
\begin{equation}
    \label{eq:orth_Y}
    \int_0^{2\pi} \int_0^{\pi} Y_{lm}(\theta,\phi)Y_{l'm'}^*(\theta,\phi) \sin{\theta}d\theta d\phi = \delta_{ll'} \delta_{mm'},
\end{equation}
where $\delta$ denotes the Kronecker delta function and
\begin{equation}
    \label{eq:orth_X}
    \int_0^{2\pi} \int_0^{\pi} \mathbf{X}_{lm}(\theta,\phi) \cdot  \mathbf{X}_{l'm'}^*(\theta,\phi) \sin{\theta}d\theta d\phi = l ( l + 1 ) \delta_{ll'} \delta_{mm'},
\end{equation}
and a simple equation that relates $\mathbf{Z}_{lm}$ with $\mathbf{X}_{lm}$, namely \cite{morse}
\begin{equation}
    \label{eq:ZX}
     \mathbf{X}_{lm} (\mathbf{r}) = -\hat{\mathbf{r}} \times \mathbf{Z}_{lm} (\mathbf{r}).
\end{equation}
We note that there is no unique convention in the definition (symbols and formal expressions) of vector spherical harmonics. For example, the classic \cite{bohr} defines our $\mathbf{M}_{lm}$ and $\mathbf{N}_{lm}$ as vector spherical harmonics. We are closer in spirit to another standard reference \cite{morse} (though note the different normalization in the definition of scalar spherical harmonics and for the symbols match $\mathbf{P}$, $\mathbf{B}$, and $\mathbf{C}$ with our $\hat{\mathbf{r}}Y$, $\mathbf{Z}$, and $\mathbf{X}$).

The implementation of the vector and scalar spherical harmonics, and by extension, of the scalar and vector wave functions, is not a simple task; for this purpose we relied on the routines of the toolbox OpenSANS \cite{sans}.

\section{Derivation of Eq. (\ref{eq:det}) with or without excitation}
\label{app:matrix}

Combining Appendix \ref{app:appvector}, together with (\ref{eq:mode}), we can reformulate the field distribution. The derivation of  (\ref{eq:det}) with or without excitation, is then a matter of application of the boundary conditions. We start with the nondriven case and repeat that these are
\begin{equation}
    \label{eq:boundcon1}
    \mathbf{n} \times \mathbf{E}(\mathbf{r}) = \mathbf{0}
\end{equation}
and
\begin{equation}
    \label{eq:boundcon2}
    \mathbf{n} \cdot \mathbf{P}_f(\mathbf{r}) = 0,
\end{equation}
on the surface $r=R$. (\ref{eq:boundcon1}) can be written, via (\ref{eq:mode}), (\ref{eq:N2}), (\ref{eq:L2}), and (\ref{eq:ZX}), as
\begin{equation}
    \label{eq:tangE1}
   - b_{lm} \frac{1}{kr} \frac{\partial \left( r j_l( k r )  \right)}{\partial r} \mathbf{X}_{lm}( \theta, \phi )
   - c_{lm} \frac{ j_l( \kappa r  ) }{ \kappa r } \mathbf{X}_{lm}( \theta, \phi ) = \mathbf{0}.
\end{equation}
Next, we multiply both sides by $\mathbf{X}_{l'm'}^{*}$, integrate along the elementary solid angle, and use (\ref{eq:orth_X}) to get
\begin{equation}
    \label{eq:tangE2}
   -  b_{lm} l ( l + 1 ) \frac{1}{kr} \frac{\partial \left( r j_l( k r )  \right)}{\partial r}
   -  c_{lm} l ( l + 1 ) \frac{ j_l( \kappa r  ) }{ \kappa r }  = 0.
\end{equation}
Since $l$ is a positive integer
\begin{equation}
    \label{eq:tangE3}
   b_{lm} \frac{1}{kr} \frac{\partial \left( r j_l( k r )  \right)}{\partial r}
   + c_{lm} \frac{ j_l( \kappa r  ) }{ \kappa r }  = 0.
\end{equation}

To deduce a similar equation from (\ref{eq:boundcon2}), we must follow a more involved course. Take (\ref{eq:pf}) in a source--free region
\begin{equation}
    \label{eq:pf_field1}
    \mathbf{P}_{f} ( \mathbf{r} ) = - \epsilon_0 \frac{\omega_p^2}{\omega(\omega + i\gamma)} \left[ \mathbf{E}( \mathbf{r} ) - \frac{\beta^2}{\omega_p^2} \nabla \left(\nabla \cdot \mathbf{E}( \mathbf{r} ) \right) \right].
\end{equation}
The Helmholtz decomposition theorem allows for writing the electric field as the superposition of a transverse and a longitudinal component, $\mathbf{E}(\mathbf{r}) = \mathbf{E}^T(\mathbf{r}) + \mathbf{E}^L(\mathbf{r})$, the first being solenoidal, the latter irrotational. Using this and the vector identity $\nabla(\nabla \cdot \mathbf{A}) = \nabla^2 \mathbf{A} + \nabla \times \nabla \times \mathbf{A}$,
\begin{equation}
    \label{eq:pf_field2}
    \mathbf{P}_{f}( \mathbf{r} ) =  - \epsilon_0  \frac{\omega_p^2}{\omega(\omega + i\gamma)} \left( \mathbf{E}^T( \mathbf{r} ) + \mathbf{E}^L( \mathbf{r} ) - \frac{\beta^2}{\omega_p^2} \nabla^2 \mathbf{E}^L( \mathbf{r} ) \right).
\end{equation}
The longitudinal component satisfies a vector Helmholtz equation $\nabla^2 \mathbf{E}^L(\mathbf{r}) + \kappa^2 \mathbf{E}^L(\mathbf{r}) = \mathbf{0}$ \cite{yuri3}. The vector Laplacian of (\ref{eq:pf_field2}) can then be eliminated
\begin{equation}
    \label{eq:pf_field3}
    \mathbf{P}_{f}( \mathbf{r} ) =  - \epsilon_0  \frac{\omega_p^2}{\omega(\omega + i\gamma)} \left( \mathbf{E}^T( \mathbf{r} ) + \mathbf{E}^L( \mathbf{r} ) + \frac{\beta^2}{\omega_p^2} \kappa^2 \mathbf{E}^L( \mathbf{r} ) \right),
\end{equation}
and after substituting (\ref{eq:kappa}) into (\ref{eq:pf_field3}) and slightly manipulating the material functions, we end up with \cite{fedor}
\begin{equation}
    \label{eq:pf_field4}
    \mathbf{P}_{f}( \mathbf{r} ) =  (\epsilon - \epsilon_0)\mathbf{E}^T( \mathbf{r} ) - \epsilon_0\mathbf{E}^L( \mathbf{r} ) = \epsilon\mathbf{E}^T( \mathbf{r} ) - \epsilon_0\mathbf{E}( \mathbf{r} ).
\end{equation}

With this more amenable format, we can return to (\ref{eq:boundcon2}) and substitute directly $\mathbf{E}^T(\mathbf{r}) = b_{lm} \mathbf{N}_l^m(\mathbf{r})$ and \mbox{$\mathbf{E}^L(\mathbf{r}) = c_{lm} \mathbf{L}_l^m(\mathbf{r})$} and yield
\begin{equation}
    \label{eq:sys2_1}
    \begin{split}
    & b_{lm} l (l + 1) (\epsilon - \epsilon_0) \frac{j_l ( k r )}{k r } Y_{lm}( \theta, \phi ) \\    
    -& c_{lm} \epsilon_0 \frac{ \partial j_l ( \kappa r )  }{ \partial( \kappa r ) } Y_{lm}( \theta, \phi ) = 0.
    \end{split}
\end{equation}
As before, multiply by $Y_{l'm'}^{*}( \theta, \phi )$, integrate along the elementary solid angle, and use (\ref{eq:orth_Y}) to get (after dividing by $\epsilon_0$)
\begin{equation}
    \label{eq:pfn7}
    b_{lm} l (l + 1) \left(\frac{\epsilon}{\epsilon_0} - 1 \right) \frac{j_l ( k r )}{k r }  - c_{lm} \frac{ \partial j_l ( \kappa r ) }{ \partial( \kappa r ) } = 0.
\end{equation}
By writing (\ref{eq:tangE3}) and (\ref{eq:pfn7}) in a matrix format, we end up with (\ref{eq:det}).

When an excitation is introduced, the procedure remains. However, the boundary conditions can be rewritten in order to distinguish clearly the ``excitation'' and ``scattered'' field, namely
\begin{equation}
    \label{eq:boundcon1_exc}
    \mathbf{n} \times \mathbf{E}_{\rm sca}(\mathbf{r}) + \mathbf{n} \times \mathbf{E}_{\rm exc}(\mathbf{r}) = \mathbf{0} 
\end{equation}
and
\begin{equation}
    \label{eq:boundcon2_exc}
    \mathbf{n} \cdot \left(\left( \epsilon - \epsilon_0 \right)\mathbf{E}^T_{\rm sca} - \epsilon_0\mathbf{E}^L_{\rm sca}\right) + \mathbf{n} \cdot ( \epsilon - \epsilon_0 )\mathbf{E}^T_{\rm exc} = 0 ,
\end{equation}
since the excitation we use here does not have a longitudinal component.

It is simple to construct the corresponding equations to (\ref{eq:tangE1}) and (\ref{eq:sys2_1})
\begin{equation}
    \label{eq:tangE1_exc}
    \begin{split}
    & - b_{lm} \frac{1}{kr} \frac{\partial \left( r j_l( k r )  \right)}{\partial r} \mathbf{X}_{lm}( \theta, \phi )
   - c_{lm} \frac{ j_l( \kappa r  ) }{ \kappa r } \mathbf{X}_{lm}( \theta, \phi ) \\
   &   - \frac{1}{kr} \frac{\partial \left( r  h_1^{(1)} ( k r )   \right)}{ \partial r} \mathbf{ X }_{10}( \theta, \phi )  = \mathbf{0}
   \end{split}
\end{equation}
and
\begin{equation}
    \label{eq:sys2_1_exc}
    \begin{split}
    & b_{lm} l (l + 1) (\epsilon - \epsilon_0) \frac{j_l ( k r )}{k r } Y_{lm}( \theta, \phi ) - c_{lm} \epsilon_0 \frac{ \partial j_l ( \kappa r )  }{ \partial( \kappa r ) } Y_{lm}( \theta, \phi ) \\
    & + 2( \epsilon - \epsilon_0) \frac{h_1^{(1)} ( k r )}{ k r } Y_{10} ( \theta, \phi ) = 0,
    \end{split}
\end{equation}
respectively. We repeat the procedure and arguments involving orthogonality, though now we use $\mathbf{X}_{1,0}^*( \theta, \phi )$ and $Y_{1,0}^{*} ( \theta, \phi )$, yielding then the correspoding to (\ref{eq:tangE3}) and (\ref{eq:pfn7})
\begin{equation}
    \label{eq:tangE3_exc}
    b_{10} \frac{1}{kr} \frac{\partial \left( r j_1( k r )  \right)}{\partial r}
   +c_{10} \frac{ j_1( \kappa r  ) }{ \kappa r } 
   = -\frac{1}{kr} \frac{\partial \left( r  h_1^{(1)} ( k r )   \right)}{ \partial r}
\end{equation}
and
\begin{equation}
    \label{eq:pfn7_exc}
    \begin{split}
    b_{10} 2 \left( \frac{\epsilon}{\epsilon_0} - 1 \right) \frac{j_1 ( k r )}{k r }  - c_{10}  \frac{ \partial j_1 ( \kappa r )  }{ \partial( \kappa r ) } 
    = -2 \left( \frac{\epsilon}{\epsilon_0} - 1 \right) \frac{h_1^{(1)} ( k r )}{ k r },
    \end{split}
\end{equation}
respectively. Cast the equations above in a matrix format to get
\begin{equation}
    \label{eq:mat_sys}
    \begin{aligned}
    \begin{bmatrix}
         \displaystyle \frac{ 1 }{ kr } \frac{\partial \left[ r j_1( k r)  \right]}{\partial r} &  \displaystyle \frac{ j_1( \kappa r )}{ \kappa r } \\
         \displaystyle 2 \left( \frac{\epsilon}{\epsilon_0} - 1 \right) \frac{ j_1( k r )}{ k r } & - \displaystyle \frac{ \partial j_1( \kappa r )}{ \partial( \kappa r ) }
    \end{bmatrix}
    \begin{bmatrix}
        b_{10} \\
        c_{10}
    \end{bmatrix} = \\ \\  
    \begin{bmatrix}
        \displaystyle -\frac{1}{kr} \frac{ \partial \left[ r h_1^{(1)}( k r )\right]}{ \partial r }  \\
        
        \displaystyle -2 \left( \frac{\epsilon}{\epsilon_0} - 1 \right) \frac{h_1^{(1)}( k r )}{k r } 
    \end{bmatrix},
    \end{aligned}
\end{equation}
at $r=R$, which is (\ref{eq:det}) driven by the excitation described by (\ref{eq:exc}).

\section{Codes for the first numerical example}
\label{app:code}
The full MATLAB code for the first numerical example is available \href{https://gitlab.kuleuven.be/u0142839/uniqueness-numerical-experiment}{here}.

\ifCLASSOPTIONcaptionsoff
  \newpage
\fi


\bibliographystyle{IEEEtran}

%



\end{document}